\documentclass[superscriptaddress,showpacs,pre]{revtex4-1}

\usepackage{graphics,amsfonts,epsfig,amsmath,verbatim,amssymb}

\def\ba{\begin{eqnarray}}
\def\ea{\end{eqnarray}}
\def\beq{\begin{eqnarray}}
\def\eeq{\end{eqnarray}}
\def\be{\begin{equation}}
\def\ee{\end{equation}}

\def\rmd{{\rm d}}
\def\rme{{\rm e}}
\def\H{{\mathcal H}}
\def\tm{{\widetilde M}}

\def\bx{{\mathbf x}}
\def\bxp{{\mathbf x'}}
\def\br{{\mathbf r}}
\def\bQ{{\mathbf Q}}
\def\bq{{\mathbf q}}

\def\H{{\cal H}} 

\def\be{\begin{equation}}
\def\ee{\end{equation}}

\def\bea{\begin{eqnarray}}
\def\eea{\end{eqnarray}}

\def\rmd{{\rm d}}
\def\rme{{\rm e}}

\newcommand{\reff}[1]{(\ref{#1})}
\newcommand\ie{{\it i.e.}}
\newcommand\eg{{\it e.g.}}
\newcommand\cf{c.f.}

\newcommand\bg{{\rm (ng)}}
\newcommand\mc{{\rm (MC)}}
\newcommand{\tmic}{t_\mu}

\newcommand{\eps}{\epsilon}

\begin{document}

\title{Dynamic crossover in the persistence probability of manifolds at criticality} 

\author{Andrea Gambassi}
\affiliation{SISSA - International School for Advanced Studies and INFN, via
  Bonomea 265, 34136 Trieste, Italy}
\author{Raja Paul}
\affiliation{Indian Association for the Cultivation of Science, Department of Solid State Physics,
Jadavpur, Kolkata 700 032, India} 
\author{Gr\'egory Schehr}
\affiliation{Laboratoire de Physique Th\'eorique (UMR du
  CNRS 8627), Universit\'e de Paris-Sud, 91405 Orsay Cedex,
  France}


\begin{abstract}
We investigate the persistence properties of critical  $d$-dimensional systems
relaxing from an initial state with non-vanishing
order parameter (\eg, the magnetization in the Ising model), 
focusing on the dynamics of the global order parameter of a $d'$-dimensional
manifold.
The persistence probability $P_c(t)$ shows three distinct long-time decays
depending on the value of the parameter $\zeta = (D-2+\eta)/z$ which also
controls the relaxation of the persistence probability in the case of a disordered initial
state (vanishing order parameter) as a function of the codimension $D = d-d'$
and of the critical exponents $\eta$ and $z$. We find that the asymptotic behavior of $P_c(t)$ is  
exponential for $\zeta > 1$, stretched exponential for $0 \leq \zeta \leq 1$, and 
algebraic for $\zeta < 0$. 
Whereas the exponential and
stretched exponential relaxations are not affected by the initial 
value of the order parameter, 
we predict and observe a crossover between two different power-law decays when
the algebraic relaxation occurs, as in the case $d'=d$ of the global order
parameter.
We confirm via Monte Carlo simulations our analytical predictions by studying  
the magnetization of a line and of a plane of the two- and three-dimensional
Ising model, respectively, with Glauber dynamics. 
The measured exponents of the ultimate algebraic decays 
are in a rather good agreement with our analytical
predictions for the Ising universality class. 
In spite of this agreement, the expected scaling behavior of the persistence probability as a function of time and of the  initial value of the order parameter remains problematic.
In this context, the non-equilibrium dynamics of the $O(n)$ model in the limit $n \to \infty$ and its subtle connection
with the spherical model is also discussed in detail. 
In particular, we show that the correlation functions of the components of the order parameter which are respectively parallel and  transverse to its average value within the $O(n\rightarrow\infty)$ model correspond to the correlation functions of the local and global order parameter of the spherical model. 
\end{abstract}

\maketitle

\section{Introduction}

After decades of research, understanding the statistic of first-passage times for 
non-Markovian stochastic processes remains a challenging issue. Of particular interest in
this context is the persistence probability $P_c(t)$, which, for a  stochastic
process $X(t\ge 0)\in{\mathbb R}$ of zero mean $\langle X(t) \rangle = 0$, is defined as the
probability that $X$ does not change sign within the time interval $[0,t]$. 
In terms of $P_c(t)$, the probability density of the first time at which the process crosses $X=0$ (zero crossing) is
$-\rmd P_c(t)/\rmd t$. This kind of first passage problems have been widely
studied by mathematicians since the early sixties
\cite{mcfadden,newell,slepian}, often inspired by engineering applications. During
the last fifteen years these problems have received a considerable attention in the
context of non-equilibrium statistical mechanics of
spatially extended systems, both theoretically
\cite{godreche_persistence,satya_review} and experimentally
\cite{persistence_exp}. 
In various relevant physical situations, ranging from coarsening dynamics to
fluctuating interfaces and polymer chains, $P_c(t)$ turns out to
decay algebraically at large times, $P_c(t) \sim t^{-\theta_c}$, where $\theta_c$
is a non trivial exponent, the prediction of which becomes particularly
challenging for non-Markovian processes~\cite{satya_review}.

In statistical physics, persistence properties were first
studied for the coarsening dynamics of ferromagnetic spin models evolving at zero temperature
$T=0$ from a random initial conditions
\cite{godreche_persistence}. In this case 
the {\it local} magnetization, \ie, the value $\pm 1$ of a
single spin, is the physically relevant stochastic
process $X(t)$ and the corresponding persistence probability turns out to decay algebraically 
at long times. 
By contrast, at any non-vanishing temperature $T>0$, the spins fluctuate very rapidly in time due to the coupling to the thermal bath and therefore the persistence probability of an individual spin decays exponentially in time. 
However, it was shown in Ref.~\cite{majumdar_critical} that 
the {\it global} magnetization, \ie, the spatial average of the local magnetization over the entire (large) sample, is characterized by a persistence probability $P_c(t)$ --- referred to as global
persistence --- which decays algebraically in time $P_c(t) \sim t^{-\theta_g}$  at temperatures $T$ below the
critical temperature $T_c$ of the model. The case corresponding to $T=T_c$,
which we shall focus on here, is particularly interesting because $\theta_g$
turns out to be a new universal exponent associated to the critical behavior of
these systems. The global persistence for critical dynamics has since been
studied in a variety of instances~\cite{glob_persist,oerding_persist,sire} (see
also Ref.~\cite{malte_global} for a recent study of the global persistence
below $T_c$). 

In order to understand how the long-time behavior of the persistence probability at fixed $T=T_c$ crosses over from the exponential of the local magnetization to the algebraic of the global one,
Majumdar and Bray~\cite{manifold} introduced and studied the persistence of the
total magnetization of a $d'$-dimensional sub manifold of a $d$-dimensional system,
with $0 \leq d' \leq d$ (a similar idea was put forward in Ref.~\cite{sire} and further studied in Ref.~\cite{bhar10}).  
The two limiting cases $d' =0$ and $d'=d$ correspond, respectively, to the
local and the global magnetization and therefore  a
crossover from an exponential ($d'=0$) to an algebraic ($d'=d$) decay is
expected in the persistence probability as $d'$ is varied from $0$ to $d$. 
Interestingly enough, it turns out
that as a function of $d'$ the persistence probability $P_c(t)$ of the manifold  
displays \emph{three} qualitatively different long-time behaviors, depending on the value of a single 
parameter $\zeta \equiv (D-2+\eta)/z$, where
$D=d-d'$ is the co-dimension of the manifold, $z$ the dynamical critical exponent and
$\eta$ the Fisher exponent which characterizes the anomalous algebraic decay of the
static two-point spatial correlation function of the spins. As a function of $\zeta$ one
finds~\cite{manifold}
\begin{eqnarray}
P_c(t) \sim 
\begin{cases}
t^{-\theta_0(d,d')} \;&\mbox{for} \quad \zeta < 0, \\
\exp{\left(-a_1 t^\zeta\right)} \;&\mbox{for} \quad 0 \leq \zeta \leq 1, \\
\exp{\left(-b_1 t \right)} \;&\mbox{for} \quad \zeta > 1 \;,
\end{cases}
\label{crossover_codim}
\end{eqnarray}
where the exponent $\theta_0(d,d')$ is a new universal exponent which depends on both
$d'$ and $d$, whereas $a_1$ and $b_1$ are non-universal constants. Interestingly enough, from the mathematical point of view, the somewhat unexpected stretched exponential behavior for $0\le
\zeta\le 1$  emerges as a consequence of a theorem due to Newell and
Rosenblatt~\cite{newell} which connects the long-time decay of the
persistence probability of a stationary process to the long-time decay of its
two-time correlation function. 
In passing, we mention that a behavior similar to Eq.~\reff{crossover_codim} was also 
predicted and observed numerically for the non-equilibrium correlation function and the global persistence of spins close to a free surface of a semi-infinite three dimensional Ising model with Glauber dynamics~\cite{pleimling_igloi}.  

The results mentioned above --- including the crossover in
Eq.~\reff{crossover_codim} --- 
were obtained for the critical coarsening of a system which is
initially prepared in a completely disordered state characterized by a vanishing initial
value $m_0=0$ of the magnetization or, in general terms, of the order parameter. Due to the absence of symmetry-breaking fields, this initial condition implies that the average $m(t)$ of the fluctuating magnetization  at time $t$ over the possible realizations of the process vanishes.   
For $m_0\neq 0$, instead, after a non-universal transient, the average
magnetization $m(t)$ grows in time as $m(t) \propto m_0 t^{\theta'_{\rm is}}$ for $t
\ll \tau_m \propto m_0^{-1/\kappa}$ whereas, for $t \gg \tau_m$,  
$m(t)$ decays algebraically to zero as $m(t) 
\propto t^{-\beta/(\nu z)}$. These different time dependences are
characterized by the universal exponents  $\theta'_{\rm is}>0$ (the so-called initial-slip
exponent~\cite{janssen_rg}) and 
\begin{equation}
\kappa = \theta'_{\rm is} + \beta/(\nu z), 
\label{eq:kappa}
\end{equation}
where 
$\beta$, $\nu$ and $z$ are the usual static and dynamic (equilibrium) critical
exponents, respectively. 
In Ref.~\cite{us_epl}, we have demonstrated that a non-vanishing value of $m_0$ results in a temporal crossover in the persistence probability $P_c(t)$ of the  thermal fluctuations $\delta m(t)$ of the fluctuating magnetization around
its average value $m(t)$, such that
\begin{eqnarray}
\label{crossover_dyn}
P_c(t) \sim 
\begin{cases}
 t^{-\theta_0}&\quad\mbox{for}\quad t_{\rm micr} \ll t \ll \tau_m \,,\\
t^{-\theta_\infty}&\quad\mbox{for}\quad t \gg \tau_m\,,
\end{cases}
\end{eqnarray}  
where $t_{\rm micr}$ is a microscopic time scale.
On the basis of a renormalization-group analysis of the dynamics of the order
parameter up to the first order in the dimensional expansion and of Monte Carlo
simulations of the two-dimensional Ising model with Glauber dynamics, we
concluded that the two exponents $\theta_0$ and $\theta_\infty$ are indeed
different, with $\theta_0 < \theta_\infty$.

In the present study,  we investigate both analytically and numerically the interplay
between the crossovers described by Eqs.~\reff{crossover_codim} and
\reff{crossover_dyn}, focusing primarily on the case of the Ising model universality class with relaxational dynamics. 
We argue that for $\zeta > 0$ the qualitative behavior
of the persistence probability in Eq.~\reff{crossover_codim}
is not affected by a non-vanishing initial magnetization.
For $\zeta < 0$, instead, we predict a crossover between two distinct
algebraic decays characterized by different exponents $\theta_0(d,d')$ and
$\theta_\infty(d,d')$, as in Eq.~\reff{crossover_dyn}. 
The conclusions of our analysis of the time dependence of the persistence probability are summarized in 
Tab.~\ref{fig_table}. 
\begin{center}
\begin{table}[h]
\label{fig_table}
\begin{tabular}{|c||c|c|}
\hline
 \quad & $t_{\rm micr}\ll t \ll \tau_m$ & $\tau_m \ll t$ \\
\hline
\hline
$\zeta < 0$ & $P_c(t) \sim t^{-\theta_0(d,d')}$ & $P_c(t)\sim t^{-\theta_\infty(d,d')}$ \tabularnewline
\hline
$0 \leq \zeta \leq 1$ & \multicolumn{2}{c|}{$P_c(t)\sim \exp{\left(-a_1 t^\zeta\right)}$} \\
\hline
$\zeta \geq 1$ & \multicolumn{2}{c|}{$P_c(t)\sim \exp{\left(-b_1 t\right)}$} \\
\hline
\end{tabular}
\caption{Time dependence of the persistence probability $P_c(t)$ of a $d'$-dimensional submanifold of a $d$-dimensional system relaxing from an initial state with non-vanishing order parameter $m_0$, as a function of the parameter $\zeta=(d-d'-2+\eta)/z$. (See the main text for details.)}
\end{table}
\end{center}
%
The rest of the paper is organized as follows: In Sec.~\ref{sec:model_scal}, 
we describe the continuous model we shall study and we present a scaling analysis which
yields the behaviors mentioned above for the persistence probability $P_c(t\gg\tau_m)$ 
at criticality, see Tab.~\ref{fig_table}. In Sec.~\ref{sec:analyt}, we present an analytic
approach to the calculation of $\theta_\infty(d,d')$ for the Ising universality class with relaxational dynamics. In particular, 
in Sec.~\ref{sec:Gaux}, we focus on the Gaussian approximation, whereas in Sec.~\ref{sec:BGaux}, we discuss the effects of non-Gaussian fluctuations. In Sec.~\ref{sec:MC},
we present the results of our numerical simulations of the Ising model with Glauber dynamics and we compare them
with the analytical predictions of Sec.~\ref{sec:analyt}. In Sec.~\ref{sec:On}, we study the persistence of
manifolds for the $O(n)$ model in the limit $n \to \infty$ and, in passing, we discuss the connection between the non-equilibrium dynamics of this model and of the spherical model. 
Our conclusions and perspectives are then presented in Sec.~\ref{sec:conc}.

\section{Model and scaling analysis}

\label{sec:model_scal}

Here we focus primarily on the Ising model on a $d$-dimensional hypercubic
lattice, evolving with Glauber dynamics at  its critical
point and we study the persistence properties of the associated order parameter  
both analytically and numerically. 
The universal aspects of the relaxation of this model are
captured by the case $n=1$ of so-called Model A~\cite{hohenberg77} 
for the $n$-component fluctuating local order parameter 
$\varphi_\alpha(\bx,t)$, 
$\alpha=1,\ldots,n$ (\eg,
the coarse-grained density of spins at point $\bx \equiv (x_1, \dots, x_d)$ in
the Ising model):
\begin{eqnarray}
\partial_t \varphi(\bx,t) = - \frac{\delta
 \H[\varphi]}{\delta \varphi(\bx,t)} + \eta(\bx,t),
\label{def_Langevin}
\end{eqnarray}
where  $\eta(\bx,t)$ is a Gaussian white noise
with $\langle \eta_\alpha(\bx,t) \rangle = 0$ and
$\langle \eta_\alpha(\bx,t) \eta_\beta(\bxp,t') \rangle = 2 T
\delta_{\alpha\beta} \delta^d(\bx-\bxp) \delta(t-t')$ 
and $T$  is the temperature of the
thermal bath. In what follows we consider the critical case $T=T_c$. In Eq.~\reff{def_Langevin}, the friction coefficient has been set to $1$ and $\H$ is the $O(n)$-symmetric
Landau-Ginzburg functional:
\begin{equation}
\label{def_O1}
\H[\varphi] = \int \rmd^d \bx \left[ \frac{1}{2}(\nabla \varphi)^2 +
 \frac{1}{2} r_0 \varphi^2 + \frac{g_0}{4!}
 (\varphi^2)^2 \right],
\end{equation} 
where $\int \rmd^d \bx \equiv \int \prod_{i=1}^d \rmd x_i$,
$r_0$ is a parameter which has to be  tuned to a  critical
value $r_{0,c}$ in order to approach the critical point at  $T=T_c$ (here $r_{0,c}
= 0$), and $g_0 > 0$ is the bare coupling constant. 

At the initial time $t = 0$, the system is assumed to be prepared in a random spin
configuration with a non-vanishing average order parameter $M_0$ along
the direction in the internal space which will be denoted by 1,
$[\varphi_\alpha(\bx,0)]_0 = M_0\delta_{\alpha 1}$ ($\alpha=1,\ldots,n$) 
and short-range correlations $[\varphi_\alpha(\bx,0) \varphi_\beta(\bxp,0)]_0 - M_0^2\delta_{\alpha 1}\delta_{\beta 1}=
\tau_0^{-1} \delta_{\alpha\beta}\delta^d
(\bx-\bxp)$, where $[\ldots]_0$ stands for the average over the distribution of the
initial configuration. It turns out that $\tau_0^{-1}$ is irrelevant in
determining the leading scaling properties~\cite{janssen_rg} 
and therefore we set $\tau_0^{-1}=0$. 
Here we first focus on the case $n=1$ of the Ising universality class, whereas in Sec.~\ref{sec:On} we shall extend our analysis to $n>1$.
The stochastic process we are interested in is the dynamics of the
fluctuations of the magnetization around its average value, \ie, 
\begin{equation}
\psi(\bx,t) = \varphi(\bx,t) - M(t)\,,\quad \mbox{where} 
\quad M(t) = \langle \varphi(\bx,t) \rangle 
\label{eq:defpsi}
\end{equation} 
is the average magnetization and $\langle\ldots\rangle$ stands for the average
over the possible realizations of the thermal noise $\eta$. Assuming the total system to be on a hypercubic lattice of large volume $L^d$, the total (fluctuating)
magnetization of the $d'$-dimensional sub manifold is given by
\begin{equation}
\tm(x_{d'+1}, \dots, x_d, t) = \frac{1}{L^{d'}} \int\! \prod_{i=1}^{d'}\rmd
x_i\, \psi(x_1,\dots, x_{d'}, x_{d'+1},\dots,x_d,t) \label{def_mag_tilde} \;,
\end{equation}
and it depends on the remaining $D=d-d'$ spatial coordinates, which shall
be denoted by the vector ${\mathbf r} = (r_1, \dots, r_D) \equiv (x_{d'+1}, \dots, x_d)$.  

The analytical calculation of the persistence probability $P_c(t)$ of the
process $\tm(\br,t)$ relies on the observation that  the field
$\tm(\br,t)$  is a Gaussian random variable for $d'>0$. Indeed, for a $d'$-dimensional manifold
of linear size $L$, $\tm(\br,t)$ is the sum of $L^{d'}$ ($d'>0$) 
 {\it local} fluctuating degrees of freedom, \eg, spins 
in a ferromagnet, which are correlated in space only across 
a {\it finite}, time-dependent and growing correlation length $\xi(t)$. In
the thermodynamic limit $L\gg \xi(t)$ the number $\sim [L/\xi(t)]^{d'}$ of effectively independent variables contributing to the sum becomes large and the central limit theorem implies
that $\tm(\br,t)$ is a Gaussian process, for which powerful tools have been
developed in order to determine the persistence
exponent~\cite{satya_review,satya_clement_persist}. 
In particular, the Gaussian nature of the process
implies that $P_c(t)$ is solely determined by the two-time correlation function
$C_{\tm}(t,t') \equiv \langle \tm(\br, t) \tm(\br, t')\rangle$.
For $t'<t \ll \tau_m$, the correlation function $C_{\tm}(t,t')$ coincides with
the one  of the case with vanishing initial magnetization $\tau_m\rightarrow\infty$, which was studied
in Ref.~\cite{manifold}. The corresponding persistence probability $P_c(t)$
displays the behavior summarized in Tab.~\ref{fig_table} for
$t\ll\tau_m$. 
However, a non-vanishing average $M_0$ of the initial order parameter affects
the behavior of $C_{\tm}(t,t')$ as soon as $t', t \sim \tau_m$. 
In the long-time regime $t > t' \gg \tau_m$, one can take advantage of the 
results presented in Ref.~\cite{cgk-06}
for the scaling behavior of the two-time correlation function of the 
Fourier transform $\psi(\bQ,t)
= \int \rmd^d \bx \, \rme^{i \bQ . \bx} \, \psi(\bx,t)$ of the local
fluctuation $\psi(\bx,t)$ of the magnetization
[see Eq.~\reff{eq:defpsi}]. Here and in what follows, the Fourier transform of $\psi(\bx,t)$ will be simply denoted by using $\bQ\equiv (Q_1, \dots, Q_d)$ as an argument of $\psi$ and we will also assume $t>t'$. 
This yields the following scaling form for the Fourier transform $\tm(\bq,t) \propto \psi(\bQ=(0,\ldots,0,q_1,\ldots,q_D),t)$ of $\tm(\br,t)$:
\begin{eqnarray}
 \langle \tm(\bq,t) \tm(-\bq,t') \rangle  = \frac{1}{q^{2-\eta}}
 \left(\frac{t}{t'} \right)^{\theta-1} f_C\left(q^z(t-t'),\frac{t}{t'}\right)
 \;,
\label{scaling_form_manifold}
\end{eqnarray}
where $\bq = (q_1, \cdots, q_D)$, $q = |\bq|$, and  $\theta = -\beta
\delta/(\nu z) = - (d+2-\eta)/(2 z)$ \cite{cgk-06}. With the prefactor
$(t/t')^{\theta-1}$ explicitly indicated, the function $f_C$ is
such that $f_C(v, u \to \infty) = f_{C, \infty}(v)$ for fixed $v$. 
The correlation function $C_{\tm}(t,t')$ 
which determines the persistence probability is then obtained from the correlation function in 
Eq.~\reff{scaling_form_manifold} by integration over momenta
\beq
C_{\tm}(t,t') \equiv \langle \tm(\br, t) \tm(\br, t')\rangle = \int  \prod_{i=1}^{D} \frac{\rmd q_i}{2 \pi}
\langle \tm(\bq,t) \tm(-\bq,t') \rangle\,.
\label{eq:FTcorr}
\eeq
The scaling form~\reff{scaling_form_manifold} has the same structure as 
the one discussed in Ref.~\cite{manifold} for $m_0 = 0$, the only differences being in the specific value the exponent $\theta$ {\it and} in the form of the scaling function $f_C(v,u)$.  
Accordingly, the argument presented in Ref.~\cite{manifold} --- which is
actually independent of these differences ---
leads to the conclusion that the persistence probability 
for $t \gg \tau_m$ behaves as:
\begin{eqnarray}
\label{crossover_codim_mag}
P_c(t) \sim 
\begin{cases}
t^{-\theta_\infty(d',d)} &\mbox{for}\quad \zeta < 0 \\
\exp{\left(-a_1 t^\zeta\right)} &\mbox{for}\quad 0 \leq \zeta \leq 1, \\
\exp{\left(-b_1 t \right)} &\mbox{for}\quad  \zeta > 1,
\end{cases}
\end{eqnarray}
where the non-universal constants $a_1$ and $b_1$ are the same as in Eq.~\reff{crossover_codim}. Indeed, in the case  $\zeta\propto D-2+\eta \ge 0$,  the integral in Eq.~\reff{eq:FTcorr} for the equal-time correlation function $C_{\tm}(t=t',t') \propto \int  \rmd^D q \;q^{-2+\eta}$ would diverge at large wavevectors in the absence of a large-wavevector cutoff $\Lambda$. Accordingly, for times large enough compared to the temporal scale $t_\Lambda\sim \Lambda^{-z}$ associated to that cutoff the integral is dominated by the quasi-equilibrium regime $t\simeq t'$ (\ie, $t,t'\gg t_\Lambda$ with arbitrary $\tau = t-t'$), which is actually independent of the initial value $m_0$ of the magnetization and is therefore identical to the case $\zeta\ge 0$ discussed in Ref.~\cite{manifold} and corresponding to $m_0=0$.  (Note that, strictly speaking, the application of the theorem by Newell and Rosenblatt~\cite{newell} to the case $0 \leq \zeta \leq 1$ actually provides the corresponding expressions in Eqs.~\reff{crossover_codim_mag} and \reff{crossover_codim} as  upper bounds, see Ref.~\cite{manifold} for details.)

The exponent $\theta_\infty(d',d)$ which describes the algebraic decay of $P_c(t)$ for   $\zeta < 0$, instead, is a new universal exponent and it is expected to differ from $\theta_0(d',d)$ in Eq.~\reff{crossover_codim}. 
Indeed, the resulting persistence exponent depends on the specific value of 
$\theta$ and on the specific form of the scaling function in
Eq.~\reff{scaling_form_manifold}, which both depend on having $t,t' \ll \tau_m$ ($m_0=0$) or $t, t' \gg \tau_m$ ($m_0\neq 0$). 

Accordingly, we shall focus below on this novel algebraic behavior emerging
for  $\zeta < 0$, which is characterized by the exponent
$\theta_\infty(d,d')$ (see Tab.~\ref{fig_table}).

\section{Analytic calculation of the persistence exponent $\theta_\infty(d,d')$}

\label{sec:analyt}

In the case we are presently interested in, $D-2+\eta \propto \zeta < 0$ and the integral in Eq.~\reff{eq:FTcorr} --- necessarily defined with a large wave-vector cutoff $\Lambda$ --- is convergent for $\Lambda \rightarrow \infty$ so that $C_\tm(t,t) = \langle \tm^2(\br,t)\rangle$ [and therefore $C_\tm(t,t')$] has a well-defined $\Lambda$-independent limit which can be taken from the outset, provided that the expression below are understood to be valid for times much larger than the time scale $t_\Lambda$ [$=t_{\rm micr}$ of Eqs.~\reff{crossover_codim} and \reff{crossover_dyn}] associated to that cutoff. 

In order to study the persistence properties it is convenient to focus on the normalized process $X(\br, t) = \tm(\br, t)/\langle \tm^2(\br, t) \rangle^{1/2}$ associated to $\tm(\br,t)$
(see, for instance, Ref.~\cite{majumdar_critical}), which is characterized by a unit variance.
The scaling form \reff{scaling_form_manifold}, together with Eq.~\reff{eq:FTcorr},  implies that the correlation function of $X(\br, t)$ has  the following scaling form for $t>t' \gg \tau_m$
\begin{eqnarray}
\label{scaling_x}
 \langle X(\br, t) X(\br, t') \rangle = \left( \frac{t}{t'} \right)^{-\mu_\infty(d,d')} F_\infty(t/t') \;,
\end{eqnarray}
where $F_\infty(x)$ is a non-constant function, with the asymptotic behavior $F_\infty(x \to \infty) \sim \mbox{const} \neq 0$ and
\begin{eqnarray}
\mu_{\infty}(d,d') &=& -(\theta - 1) + \frac{\zeta}{2} \label{eq:hyp_a}\\
&=& 1 + \frac{d+D}{2z} \;.
\label{eq:hyp_b}
\end{eqnarray}
Even though the Gaussian process described by Eq.~\reff{scaling_x} is not stationary, it becomes such when the logarithmic time
$T = \log{t}$ is introduced. In fact, this yields $\langle X(T) X(T') \rangle = \rme^{-\mu_\infty(T-T')}F_\infty(\rme^{T-T'}) = C_{\rm st}(T-T')$ with  $C_{\rm st}(\Delta T) \sim \exp [-\mu_\infty(d,d') \Delta T]$ for large $\Delta T$, which implies \cite{slepian} that the persistence probability $P_c$ decays exponentially in logarithmic time, $P_c(T) \sim \exp{[-\theta_\infty(d,d')T]}$, \ie, an algebraic decay  
$P_c(t) \sim t^{-\theta_\infty(d,d')}$ in terms of the original time variable.  In addition, if $F_\infty(x)$ is actually independent of $x$, \ie, if it is a constant, the process can be mapped onto a Markovian process for which $\theta_\infty(d,d') = \mu_\infty(d,d')$. In a sense, $\mu_\infty$ provides a sort of "Markovian approximation" of the exponent $\theta_\infty$. In the generic case, however,  the exponent $\theta_\infty(d,d')$ depends in a non-trivial fashion on both the exponent $\mu_\infty(d,d')$ [see Eq.~\reff{scaling_x}] and the full scaling function $F_\infty(x)$, which, for the present case $D\neq 0$,  is known only within the Gaussian approximation of Eqs.~\reff{def_Langevin} and \reff{def_O1} \cite{cgk-06} (whereas for $D=0$, an expression for the first correction in the $\epsilon$-expansion about the spatial dimension $d=4-\epsilon$ is available~\cite{cgk-06}).  
It actually turns out (see further below) that the function $F_\infty(x)$ is non-trivial already within the Gaussian approximation, a fact that makes the calculation of the exponent $\theta_\infty(d,d')$ a rather difficult task. 
Here we present a perturbative expansion of this exponent which is valid for small co-dimension $D \ll 1$.
Indeed,  one notes that $F_\infty(x)$ reduces to a constant for $D=0$ and therefore the  persistence exponent  
$\theta_\infty(d,d')$ is given by its Markovian approximation $\theta_\infty(d,d') = \mu_\infty$~\cite{satya_review,us_epl}. 
For small $D$ one can take advantage of the perturbative formula derived in 
Refs.~\cite{majumdar_critical,oerding_persist} in order to expand the non-Markovian process for $D\neq 0$ and its persistence exponent $\theta_\infty$  around the Markovian one corresponding to $D=0$.

\subsection{The Gaussian approximation}

\label{sec:Gaux}

The Gaussian approximation discussed below is actually exact in
dimension $d>4$ and it is obtained by neglecting non-linear contributions
in $\psi$  to the Langevin equation~\reff{def_Langevin} once it has been expressed in terms of 
$\psi(x,t)$ and $m^2 \equiv g_0 M^2/2$~(see, \eg, Ref.~\cite{cgk-06}). This yields the evolution equation
\begin{equation}\begin{split}
&\partial_t \psi(\bx,t) = [\nabla^2  - m^2(t)] \psi(\bx,t) + \eta(\bx,t)
\\
&\mbox{where} \quad \partial_t m(t) + \frac{1}{3} m^3(t) = 0  \,,
\label{eq_gaussian}
\end{split}
\end{equation}
with the initial condition $m(t=0)=m_0$.
In order to determine the persistence exponent, we need to calculate the correlation function $C_\tm$ of the order parameter of the manifold  [see Eqs.~\reff{def_mag_tilde} and \reff{eq:FTcorr}]. In turn, $C_\tm$ can be inferred from  the two-time correlation function of the Fourier components $\psi(\bQ,t) = \int \rmd^d \bx \,\rme^{i \bQ . \bx} \psi(\bx,t)$, where $\bQ \equiv (Q_1, \dots, Q_d)$, which was calculated  in Ref.~\cite{cgk-06} (see Eq.~(59) therein): 
\be
\label{correl_fourier}
C_\bQ(t,t') \equiv \langle \psi(\bQ,t) \psi(-\bQ,t) \rangle = 
\frac{2\rme^{-\bQ^2(t+t')}}{[(t+\tau_m)(t'+\tau_m)]^{3/2}}
\int_0^{t_<}\rmd t_1 (t_1 + \tau_m)^3 \rme^{2 \bQ^2 t_1} \;,
\ee
where $t_< = {\rm min}\{t,t'\}$ and $\tau_m = 3/(2m_0^2)$. The correlation function $C_{\tm}(t,t') = \langle \tm(\br, t) \tm(\br, t')\rangle$ in space follows from the integration \reff{eq:FTcorr} of the correlation of $\tm(\bq,t) \propto \psi(\bQ=(0,\ldots,0,q_1,\ldots,q_D),t)$ [see Eq.~\reff{def_mag_tilde}]: 
\begin{equation}
\begin{split}
C_\tm(t,t') =& \langle \tm(\br, t) \tm(\br, t') \rangle = \int \prod_{i=1}^{D} \frac{\rmd q_i}{2 \pi} C_{\bQ = (0,\bq)}(t,t') \; \\
=& \frac{2 c_D \tau_m^{1-D/2}}{[(\tilde t+1) (\tilde{t'}+1)]^{3/2}} \int_0^{\tilde{t'}}\rmd \tilde t_1\; (\tilde t_1+1)^3 (\tilde t+ \tilde{t'} - 2 \tilde t_1)^{-D/2} \;,
\end{split}
\label{eq:cmt}
\end{equation}
where, on the first line, we used the notation  $\bQ = (0,\bq) \equiv (0,\cdots,0,q_1, q_2,\cdots, q_{D})$ and, on the second, we introduced $c_D = (4\pi)^{-D/2}$ and the dimensionless time variables $\tilde t\equiv t/\tau_m$ and ${\tilde t'}\equiv t'/\tau_m$, assuming $t'<t$.
The correlation function~$\langle X(\br, t>t') X(\br, t') \rangle$ of the normalized process $X(\br, t) = \tm(\br, t)/\langle \tm^2(\br, t) \rangle^{1/2}$ is therefore given by
\begin{equation}
\label{expr_gauss_complete}
\langle X(\br, t) X(\br, t') \rangle = 
\frac{C_\tm(t,t')}{\sqrt{C_\tm(t,t)C_\tm(t',t')}} =\left(\frac{\tilde t}{\tilde t'}\right)^{-(D+2)/4} \frac{I_{D/2}(t'/t,\tilde{t'})}{[I_{D/2}(1,\tilde{t}) I_{D/2}(1,\tilde{t'})]^{1/2}} ,
\end{equation}
where
\begin{equation}
\label{expr_gauss_complete_b}
I_{a}(x,u) = \int_0^1 \rmd v (1+u v)^3 [1+x(1-2v)]^{-a} \;.
\end{equation}
Note that in order for $C_\tm(t,t)$ to be defined, $I_{D/2}(1,u)\sim \int^1\rmd v (1-v)^{-D/2}$ has to be convergent, which requires $D<2$, consistently with the assumption $\zeta = (D-2)/2 <0$ ($\eta=0$ and $z=2$ within the present approximation).
In contrast to the case $D=0$ of the global persistence, discussed in Ref.~\cite{us_epl} [see Eq.~(6) therein], there is no suitable choice of a function $L(t)$ such that the correlation function~\reff{expr_gauss_complete} for $D\neq 0$ takes the form of a ratio $L(\tilde t')/L(\tilde t)$, which corresponds to a Markovian process.  Accordingly, the Gaussian process $\{X(t)\}_{t\ge 0}$ for $D\neq 0$ displays significant non-Markovian features even within the Gaussian approximation. 

\begin{figure}
\begin{tabular}{ccc}
\includegraphics[scale=0.6]{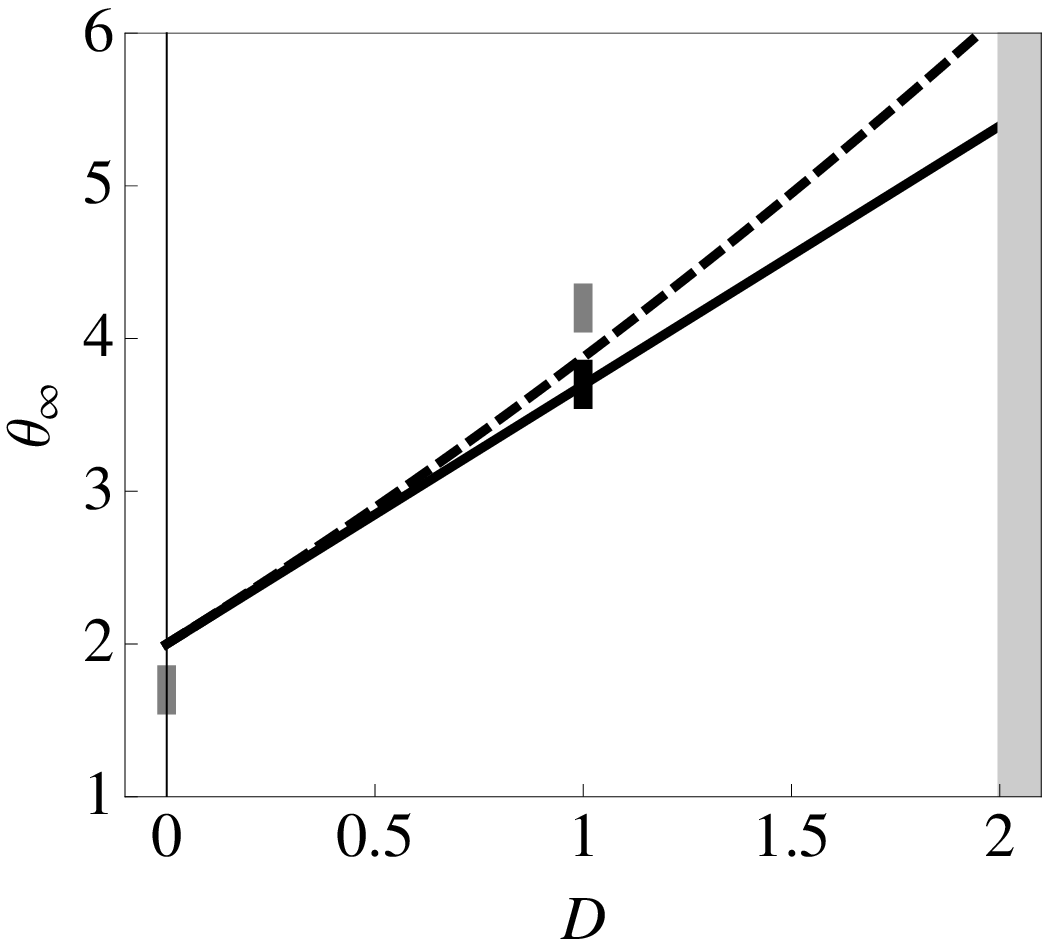}
&\quad\quad\quad&
\includegraphics[scale=0.6]{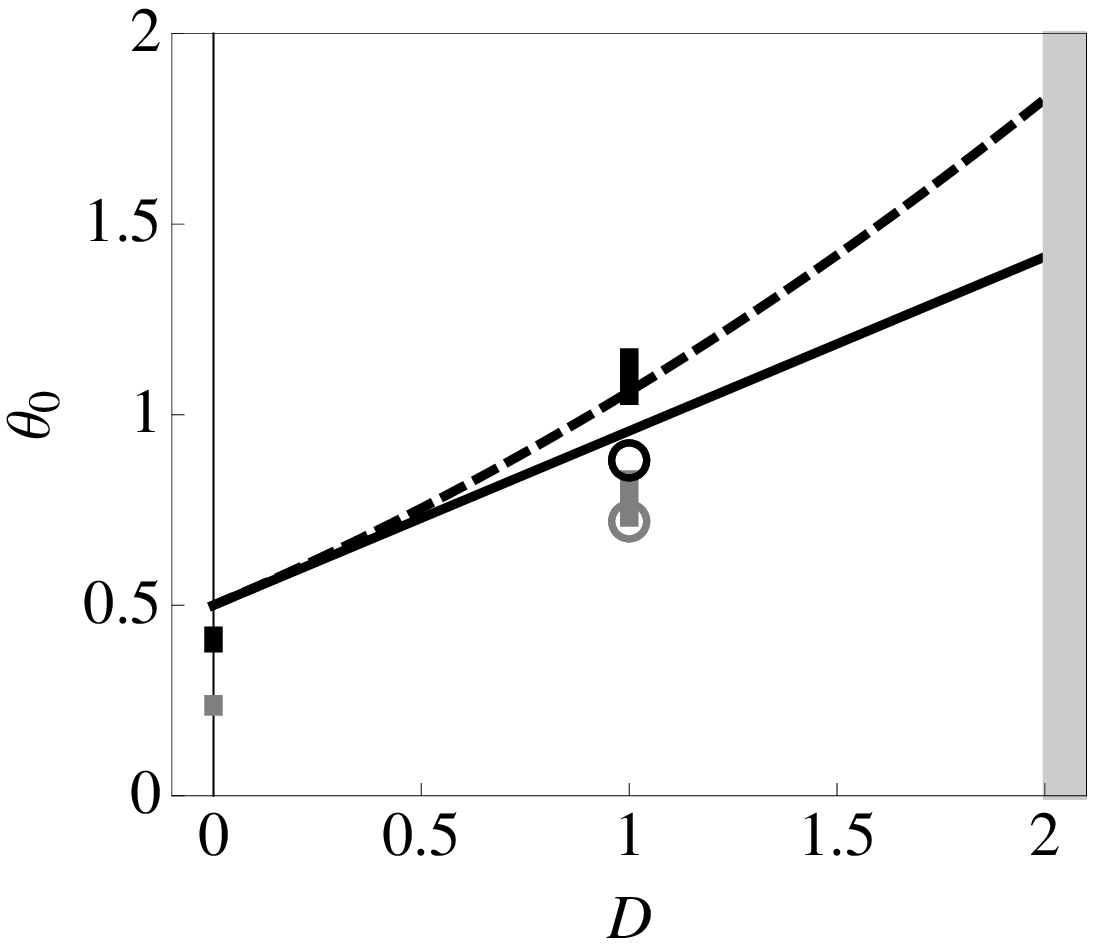}
\\[2mm]
(a)&& (b)
\end{tabular}
\caption{Persistence exponents  $\theta_{\infty,0}(d,d-D)$ of  the global order parameter of a manifold of codimension $D$ as a function of $D$, in the case of (a) non-vanishing and (b) vanishing  initial value of the order parameter. 
The solid and the dashed curves in the two panels correspond to the estimates $\theta^{[1]}_{\infty,0}$  [Eqs.~\reff{eq:est1} and~\reff{eq:est1_0}] and $\theta^{[2]}_{\infty,0}$ [Eqs.~\reff{eq:est2} and~\reff{eq:est2_0}], respectively, derived within the Gaussian model universality class with relaxational dynamics [Eq.~\reff{def_Langevin}].
For $D\ge 2$, \ie, $\zeta\ge0$ (shaded areas) the asymptotic decay of the persistence probability of this model is no longer algebraic. 
The vertical bars for $D=0$ (see Ref.~\cite{us_epl} and Refs.~\cite{Schulke97,stauffer96} for (a) and (b), respectively) and $D=1$ (present work) indicate the corresponding Monte Carlo estimates of the persistence exponents for the Ising model universality class with Glauber dynamics in $d=3$ (black) and $d=2$ (grey). On panel (b), the black and grey circles for $D=1$ indicate the preliminary Monte Carlo estimates of Ref.~\cite{manifold}.
%
%
\label{fig:Gaux}}
\end{figure}

In the asymptotic regime  $\tilde t > \tilde{t'} \gg 1$ we are interested in $I_a(x,u\gg1 ) \simeq u^3\int_0^1\rmd v \, v^3[1+x(1-2v)]^{-a}$ and therefore the correlation 
function~\reff{expr_gauss_complete} takes the form
\begin{equation}
\label{x_gen_crossover_gauss}
\langle X(\br, t) X(\br, t') \rangle \simeq (t/t')^{-(2+D/4)} {\cal A}(t'/t)\,,\quad\mbox{for}\quad 
t>t' \gg \tau_m \;,
\end{equation}
where 
\begin{equation}
\label{scaling_gaussian}
{\cal A}(x) = A_D \int_0^1 \rmd v \, v^3 [1+x(1-2v)]^{-D/2} \;,
\end{equation}
and $A_D = [2^{-D/2} \times 6 \Gamma(1-D/2)/\Gamma(5-D/2)]^{-1} = {(8-D)(6-D)(4-D)(2-D)}/(3 \times 2^{5-D/2})$, with finite ${\cal A}(0)=A_D/4$ and, by definition, ${\cal A}(1)=1$. 
The correlation function~\reff{x_gen_crossover_gauss} has the form~\reff{scaling_x} with $d = 4$, $z = 2$, \ie, $\mu_\infty = 2 + D/4$ and $F_\infty(t/t') = {\cal A}(t'/t)$. 
The scaling function ${\cal A}(t'/t)$ is indeed a non-trivial function of its argument and it reduces to a constant only for $D=0$, \ie, for vanishing codimension. As anticipated, one can take advantage of this fact in order to determine  $\theta_\infty(d,d')$ perturbatively, by expanding around the Markovian Gaussian process for $D=0$ according to 
Ref.~\cite{oerding_persist}. First, one introduces the logarithmic time $T = \log{t}$ and expands $\langle X(\br, T) X(\br, T')\rangle$ for small $D$:
\begin{equation}
 \langle X(\br, T) X(\br, T') \rangle = \rme^{-\bar \mu_\infty(T-T')} \bar{\cal A}(\rme^{-{(T-T')}}) \,, 
\end{equation} 
with $\bar\mu_\infty =2$,
\begin{equation} 
\bar{\cal A}(x) \equiv x^{D/4} {\cal A}(x)= 1 + D \bar {\cal A}_1(x) + {\cal O}(D^2) \,,
\end{equation}
and
\begin{eqnarray}
\bar{\cal A}_1(x) = \frac{1}{2} \left[-\frac{11}{6} + \frac{y}{3} + \frac{y^2}{2} + y^3 - \frac{1}{2} \log{(2y-1)} + (y^4-1) \log{(y-1) - y^4 \log{y}} \right], \;\mbox{where}\;  y = (x^{-1}+1)/2 \,.
\label{eq:cA1gaux}
\end{eqnarray}
The function $\bar{\cal A}_1(x)$ is responsible, at the lowest order in $D$,  for the non-Markovian corrections to $\theta_\infty \equiv \bar{\mathcal R} \bar\mu_\infty$ which can be calculated using the perturbation theory of Ref.~\cite{oerding_persist}: 
 \begin{equation}
\bar{\mathcal R} \equiv \frac{\theta_\infty}{\bar\mu_\infty} = 1 - D
\frac{2 \bar\mu_\infty}{\pi} \int_0^1 \!\rmd x \frac{x^{\bar\mu_\infty-1} \bar{\cal
    A}_1(x)}{(1 -  x^{2\bar\mu_\infty})^{3/2}} 
+ {\cal O}(D^2)\,. \label{eq:R}
\end{equation}
With the explicit expression for $\bar{\cal A}_1(x)$ given in Eq.~\reff{eq:cA1gaux}, a straightforward  numerical integration yields 
\begin{equation}
\bar{\mathcal R}  = 1 + D \times 0.848\cdots+ {\cal O}(D^2)\,,
 \label{theta_smallD}
\end{equation}
which renders\begin{equation}
\theta_\infty^{[1]}(d,d'=d-D) = 2 [ 1 + D \times 0.848\cdots+ {\cal O}(D^2)] \;
\label{eq:est1}
\end{equation}
as a Gaussian estimate of $\theta_\infty$.
In particular, for $D=1$ --- the case numerically studied in Sec.~\ref{sec:MC} below --- it takes the value
\begin{equation}
\theta_\infty^{[1]}(d,d'=d-1) = 3.69 \ldots \;.
\label{theta_inf_G_1}
\end{equation} 
Alternatively, a different (yet equivalent) numerical estimate of $\theta_\infty$ can be obtained by expanding  up to first order in $D$  only ${\mathcal A}(x) = 1 + D {\mathcal A}_1(x) + {\cal O}(D^2)$ on the rhs of Eq.~\reff{x_gen_crossover_gauss} while keeping the full $D$-dependence of the value of $\mu_\infty =2+D/4$, which corresponds to the Markovian approximation. Accordingly, one has $\theta_\infty(d,d') = \mu_\infty {\mathcal R} = (2+D/4) {\mathcal R}$ where the ratio ${\mathcal R}$ is given here by Eq.~\reff{eq:R} in which $\bar\mu_\infty \mapsto \mu_\infty = 2 + O(D)$ and $\bar{\mathcal A}_1(x) \mapsto {\mathcal A}_1(x) \equiv  \bar{\mathcal A}_1(x) - (\ln x)/4$. 
With these substitutions one finds ${\mathcal R} = \bar {\mathcal R} - D/8 + {\cal O}(D^2)$ and therefore
\begin{equation}
\theta_\infty^{[2]}(d,d'=d-D) = 2 (1 + D/8) [ 1 + D \times 0.723\cdots+ {\cal O}(D^2)],
\label{eq:est2}
\end{equation}
which, as expected, has the same series expansion as $\theta_\infty^{[1]}(d,d'=d-D)$ up to the first order in $D$ and provides the numerical estimate
\begin{equation}
\theta_\infty^{[2]}(d,d'=d-1) =  3.87\ldots
\label{theta_inf_G_2}
\end{equation}
for $D=1$.
These expressions for the persistence exponent $\theta_\infty(d,d-D)$ have been derived accounting only for the effects of Gaussian fluctuations of the order parameter and therefore they are increasingly accurate as the non-Gaussian fluctuations become less relevant, \ie, as the spatial dimensionality $d$ of the model approaches and exceeds 4. Accordingly, the numerical estimates $\theta_\infty^{[1,2]}(d,d-D)$ in Eqs.~\reff{eq:est1} and~\reff{eq:est2} are expected to be increasingly accurate as $d$ increases for a fixed small codimension $D$. In the next section we shall compare these analytical predictions, extrapolated to $D=1$ and provided by Eqs.~\reff{theta_inf_G_1} and~\reff{theta_inf_G_2},  to the results of Monte Carlo simulations of the Ising model with Glauber dynamics in spatial dimensionality $d=2$ and $3$, discussed in Sec.~\ref{sec:MC}. 
Figure~\ref{fig:Gaux}(a) summarizes the available estimates of $\theta_\infty(d,d-D)$ as a function of the codimension $D$ and of the space dimensionality $d$ of the model. The solid and the dashed lines correspond to the estimates $\theta^{[1]}_\infty(d,d-D)$ [Eq.~\reff{eq:est1}] and $\theta_\infty^{[2]}(d,d-D)$ [Eq.~\reff{eq:est2}], respectively, derived within the Gaussian model. The vertical bars for $D=0$ and $D=1$ indicate the corresponding Monte Carlo estimates (see Ref.~\cite{us_epl}  and Sec.~\ref{sec:MC} below, respectively) for the Ising model with Glauber dynamics in $d=2$ (grey) and $d=3$ (black). As expected, the Gaussian approximation provides a rather accurate estimate of the actual  value of $\theta_\infty$ in $d=3$. 
%
%
%
%
%
For comparison we report here also the codimension expansion of $\theta_0(d,d')$ --- corresponding to the case $m_0=0$ --- which can be determined  as detailed above for $\theta_\infty$ on the basis of Eqs.~\reff{expr_gauss_complete} and 
\reff{expr_gauss_complete_b} in the limit $\tilde{t'} < \tilde t \ll 1$. In this case, $I_a(x,u\ll 1)\simeq \int_0^1\rmd v [1+x(1-2v)]^{-a} = [(1+x)^{1-a}-(1-x)^{1-a}]/[2(1-a)x]$ and therefore
\begin{equation}
\label{x_gen_crossover_gauss_0}
\langle X(\br, t) X(\br, t') \rangle \sim (t/t')^{-(2+D)/4} {\cal B}(t'/t)\,,\quad\mbox{for}\quad 
\tau_m \gg t>t'\;,
\end{equation}
where 
\begin{equation}
{\cal B}(x) = \frac{(1+x)^{1-D/2}-(1-x)^{1-D/2}}{2^{1-D/2}x} \;,
\label{scaling_gaussian_0}
\end{equation}
with finite ${\cal B}(0)=2^{D/2}(1-D/2)$ and, by its very definition,  ${\cal B}(1)=1$. The Markovian approximation to $\theta_0$ is given by $\mu_0 \equiv 1/2+D/4$ and corresponds to ${\cal B}(x)\equiv 1$ in Eq.~\reff{x_gen_crossover_gauss_0}, \ie, to having $D=0$. 
This expression for $\mu_0$ agrees with the hyperscaling relation~\cite{majumdar_critical}
\begin{equation}
\mu_0 = -(\theta_{\rm is}-1) + \zeta/2 \;,
\label{eq:hyp0}
\end{equation}
which can be derived by the same arguments which lead to Eq.~\reff{eq:hyp_a}, taking into account that in this case $m_0=0$ the correlation function of the order parameter fluctuations  scales as in Eq.~\reff{scaling_form_manifold} with $\theta$ replaced by the initial-slip exponent 
$\theta_{\rm is} \equiv \theta_{\rm is}'-(2-\eta-z)/z$~\cite{janssen_rg}.  For the present Gaussian model $\theta_{\rm is}=0$ and $\eta=0$. 
(Note that with the identification $\lambda_c/z\mapsto -(\theta_{\rm is}-1)+\zeta$, Eq.~\reff{eq:hyp0} renders the expression for the persistence exponent $\mu_0$ which is reported  after Eq.~(8) in Ref.~\cite{manifold} in terms of $\lambda_c$.) 
Taking advantage of Eq.~\reff{eq:R} and the series expansion of ${\mathcal B}$ for small $D$ one finds
\begin{equation}
\theta_0^{[1]}(d,d-D) = \frac{1}{2} [1 + D \times (\sqrt{2}-1/2) + {\cal O}(D^2)] \;,
\label{eq:est1_0}
\end{equation}
which coincides with the case $d=4$ of the expression derived in Ref.~\cite{manifold} for the $O(n\rightarrow\infty)$ model.  An equivalent estimate of $\theta_0$ is obtained as discussed above for $\theta_\infty$: 
\begin{equation}
\theta_0^{[2]}(d,d-D) = \frac{1}{2}(1+D/2)[1 + D \times (\sqrt{2}-1) + {\cal O}(D^2)] \;.
\label{eq:est2_0}
\end{equation}
Figure~\ref{fig:Gaux}(b) summarizes the available estimates of $\theta_0(d,d-D)$ as a function of the codimension $D$ and of the space dimensionality $d$ of the model. As in panel (a) of the same figure, the solid and the dashed lines correspond to the estimates $\theta^{[1]}_0(d,d-D)$ [Eq.~\reff{eq:est1_0}] and $\theta_0^{[2]}(d,d-D)$ [Eq.~\reff{eq:est2_0}], respectively, derived within the Gaussian model. The vertical bars for $D=0$ and $D=1$ indicate the corresponding Monte Carlo estimates (see Refs.~\cite{Schulke97,stauffer96}  and Sec.~\ref{sec:MC} below, respectively) for the Ising model with Glauber dynamics in $d=2$ (grey) and $d=3$ (black). On panel (b), the gray and black circles for $D=1$ indicate the corresponding preliminary Monte Carlo estimates reported in Ref.~\cite{manifold}, which turn out to be, respectively, marginally compatible and significantly different from the ones in $d=2$ (grey) and $d=3$ (black) presented below in Sec.~\ref{sec:MC}.

\subsection{Beyond the Gaussian approximation (perturbatively)}

\label{sec:BGaux}

According to the analytical predictions within the Gaussian model indicated by the solid and dashed lines in Fig.~\ref{fig:Gaux}, the persistence exponents $\theta_{\infty,0}^{[1,2]}(d,d-D)$ \emph{increase} upon increasing the codimension $D\ge 0$, until $D$ reaches the value $D=2$, above which (shaded areas in Fig.~\ref{fig:Gaux}) the relaxation of the persistence probability is no longer algebraic. 
The contributions of non-Gaussian fluctuations to the global persistence exponents $\theta_\infty$  and $\theta_0$ within the $O(n)$ universality class have been  calculated analytically and perturbatively  
in Refs.~\cite{us_epl} and~\cite{majumdar_critical,oerding_persist}, respectively, only for the case $D=0$. (For $D>0$ the persistence exponent $\theta_0$ is studied in Ref.~\cite{manifold} only in the limit $n\rightarrow\infty$, see Sec.~\ref{sec:On} below.)
For the Ising universality class ($n=1$) these corrections  \emph{decrease} $\theta_{\infty,0}$ compared to its Gaussian value, as the spatial dimensionality $d$ of the model decreases below the upper critical dimensionality $d=4$.  
On the basis of these behaviors of $\theta_{\infty,0}(d,d-D)$ for $d>4$ (Gaussian model) as a function of (small) $D>0$ and for $D=0$ as a function of (small) $4-d$, it is difficult to predict analytically the qualitative dependence on $D$ of  $\theta_{\infty,0}(d,d-D)$ at fixed $d$ resulting from the combined effect of these two competing trends as a function of $D$ and $d$. The same problem arises in the more general case of the $O(n)$ universality class with generic $n$. %
In this respect, it would be desirable to extend to the case $D\neq 0$ the analysis of the contribution of non-Gaussian fluctuations to the persistence exponent of the $O(n)$ universality class, beyond the $n\rightarrow\infty$ limit, following
Ref.~\cite{us_epl} and Refs.~\cite{majumdar_critical,oerding_persist} for the cases $m_0\neq 0$ and $m_0=0$, respectively.
This  requires the knowledge of the analytic expression (\eg, in a dimensional expansion around the upper critical dimensionality $d=4$) of the wave-vector dependence of the correlation function of the order parameter $C_\bQ(t,t')$ 
beyond the Gaussian approximation [Eq.~\reff{correl_fourier}]. At present, however, such an analytic expression within the $O(n)$ universality class with $m_0\neq 0$ is available only for $\bQ=0$ (see Refs.~\cite{cgk-06,andrea_ordered_on}) and,  in the limit $n\rightarrow \infty$, for transverse fluctuations (see Ref.~\cite{andrea_ordered_on}; in Sec.~\ref{sec:sp-On} we shall comment on the relation between the $O(n\rightarrow\infty)$ model and the spherical model investigated, in this context, in Refs.~\cite{as-06,as-new}). For a vanishing initial value of the order parameter $m_0=0$, instead, the expression of $C_\bQ(t,t')$ beyond the Gaussian approximation and for finite $n$ is known for generic 
$\bQ$ only up to the first order in the dimensional expansion around $d=4$~\cite{cg-02a} and up to the second order for $\bQ=0$~\cite{cg-02b} (see Ref.~\cite{critical_review} for a review). 
However, we point out here that in order to observe a non-trivial interplay between the effects of a non-vanishing codimension $D\neq 0$ and those of non-Gaussian fluctuations for $d<4$  one would need to account for higher-order terms in the expansion of ${\mathcal R}$ around a Markovian process, unless the dependence of the correlation function $C_\bQ(t,t')$ on the dimensionality $d$  is known non-perturbatively as in the case $n\rightarrow\infty$ discussed in Ref.~\cite{manifold} and in Sec.~\ref{sec:On} below. 
Indeed, the general structure of the correlation function of the normalized process is 
$\langle X(t)X(t')\rangle = (t/t')^{-\mu(d,D)} F(t'/t;d ,D)$ 
with $t>t'$, where the exponent $\mu(d,D)$ and the function $F$ depend on the specific process under study and $\mu(d,D)$ is determined such that $F(0;d,D)$ is finite and non-zero [note that, by definition, $F(1;d,D)=1$]. If $F(x;d,D)$ turns out to be independent of $x$ [and therefore $F(x;d,D)\equiv 1$], then the process is Markovian with persistence exponent $\mu(d,D)$ which, in turn,  is related to known exponents via hyperscaling relations such as Eqs.~\reff{eq:hyp_a} and \reff{eq:hyp0}.
In the limit $n\rightarrow\infty$ this is actually the case for $D=0$ and generic $d$ (c.f. Sec.~\ref{sec:On}), \ie, $F(x;d,D=0)\equiv 1$. Accordingly, the expansion for small $D$ of $F(x;d,D)$ takes the form $F(x;d,D) = 1 + D \partial_DF(x;d,0) + {\cal O}(D^2)$ and one can use the formula presented in Refs.~\cite{majumdar_critical,oerding_persist} [see Eq.~\reff{eq:R} above]  in order to calculate the correction ${\mathcal R}(d,D) =  1 + D I_d + {\cal O}(D^2)$ to $\mu(d,D)$ which determines the persistence exponent $\theta(d,D) \equiv \mu(d,D)\times{\mathcal R}(d,D)$. Here the deviation from the Markovian evolution is controlled in terms of the (small) parameter $D$ and ${\mathcal R}$ has a non-trivial dependence on the dimensionality $d$ of the model. This approach was adopted in Ref.~\cite{manifold} for $m_0=0$ and, c.f., in Sec.~\ref{sec:On} of the present work for $m_0\neq 0$. In the case of finite $n$, instead, the correlation function $C_\bQ(t,t')$ for $d<4$ is typically known in a dimensional expansion around $d=4$, up to a certain order in $\eps = 4 - d$. As a consequence, $F(x;d,D) = F(x;4,D) - \eps \partial_dF(x;4,D) + {\cal O}(\eps^2)$, where the lowest-order term $F(x;4,D)$ corresponds to the Gaussian approximation, discussed above for $n=1$.  
[The line of argument presented below actually applies also to those cases in which the first correction term to $C_\bQ(t,t')$ and therefore to $F(x;d,D=0)$ is of ${\cal O}(\eps^2)$~\cite{majumdar_critical,oerding_persist}.]
In the generic case, $F(x;4,D)$ as a function of $x$ is not constant, resulting in a non-Markovian process even for $\eps=0$. However, the process turns out to be Markovian for $D=0$, \ie, $F(x;d=4,D=0) \equiv 1$, and therefore one is naturally led to perform a codimension expansion of   $F(x;4,D) = 1 + D \partial_DF(x;4,0) + {\cal O}(D^2)$ and, for consistency, of 
$\partial_dF(x;4,D) = \partial_dF(x;4,0)  + {\cal O}(D)$. Accordingly, one has
$F(x;d,D) = 1 + D \partial_DF(x;4,0) - \eps  \partial_dF(x;4,0) + {\cal O}(D^2,\eps^2,\eps D)$, where the deviations from the non-Markovian evolution are jointly controlled by $D$ and $\eps$. Due to the fact that the perturbative expression for ${\mathcal R}$ [see, \eg, Eq.~\reff{eq:R}] is valid up to the first order in the deviation from the Markovian evolution, all the terms of ${\cal O}(D^2,\eps^2,\eps D)$ in $F(x;d,D)$ can be neglected when calculating 
\begin{equation}
{\mathcal R}(d=4-\eps,D) \equiv 1 + D\, A + \eps\, B+ {\cal O}(\eps^2,\eps D,D^2),
\label{eq:RepsD}
\end{equation}
where the coefficients $A$ and $B$ are given by the integrals of $\partial_DF(x;4,0)$ and $-\partial_dF(x;4,0)$, respectively, according to the rhs of Eq.~\reff{eq:R} in which, for consistency, the value $\mu(d=4,D=0)$ of $\mu$ at the lowest order in $\eps$ and $D$ enters. 
Due to the double expansion in Eq.~\reff{eq:RepsD}, the lowest-order correction to $\mu(d,D)$ which results from both a finite codimension $D$ and non-Gaussian fluctuations for $d<4$ is given by the superposition of the two corresponding corrections taken separately, which  --- for the $O(n)$ universality class --- can be inferred from Ref.~\cite{manifold} and the present study ($D>0$, $\eps =0$), and from 
Refs.~\cite{majumdar_critical,oerding_persist,us_epl} ($D=0$, $\eps >0$), respectively.

\begin{figure}
\begin{tabular}{ccc}
\includegraphics[scale=0.6]{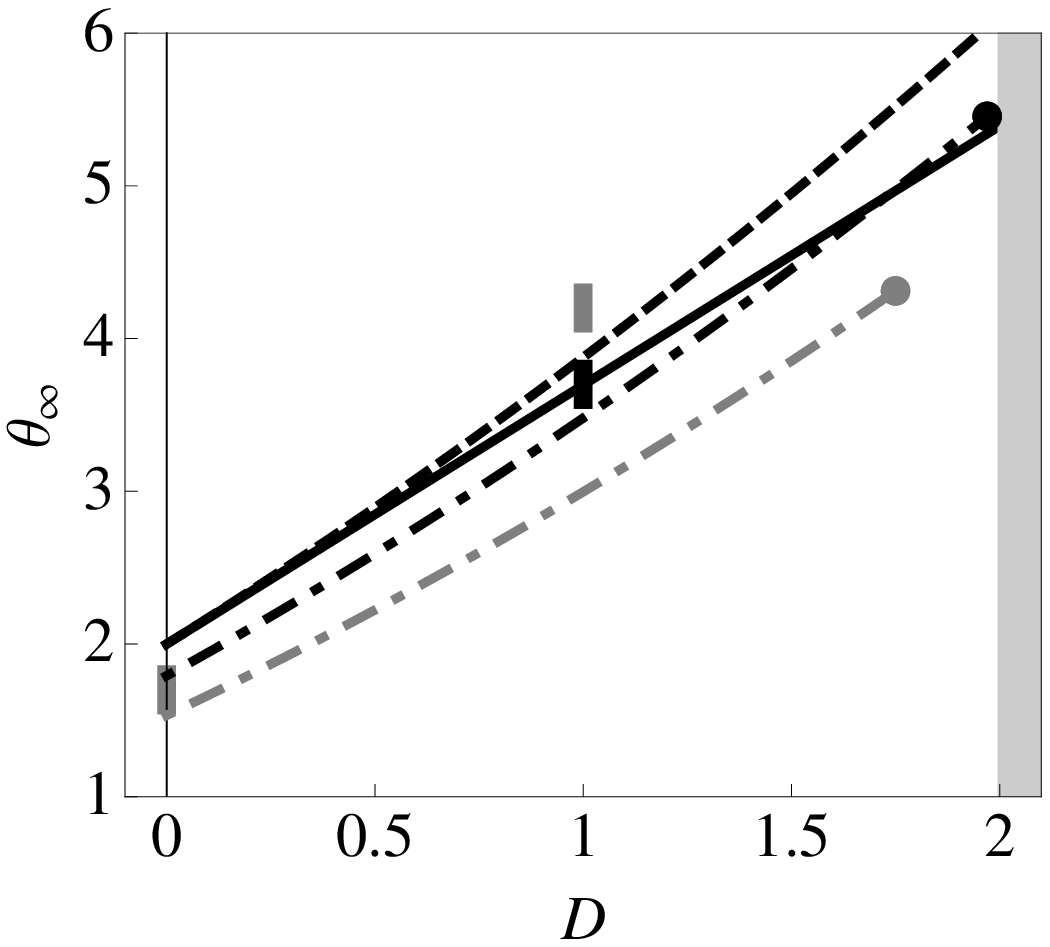}
&\quad\quad\quad&
\includegraphics[scale=0.6]{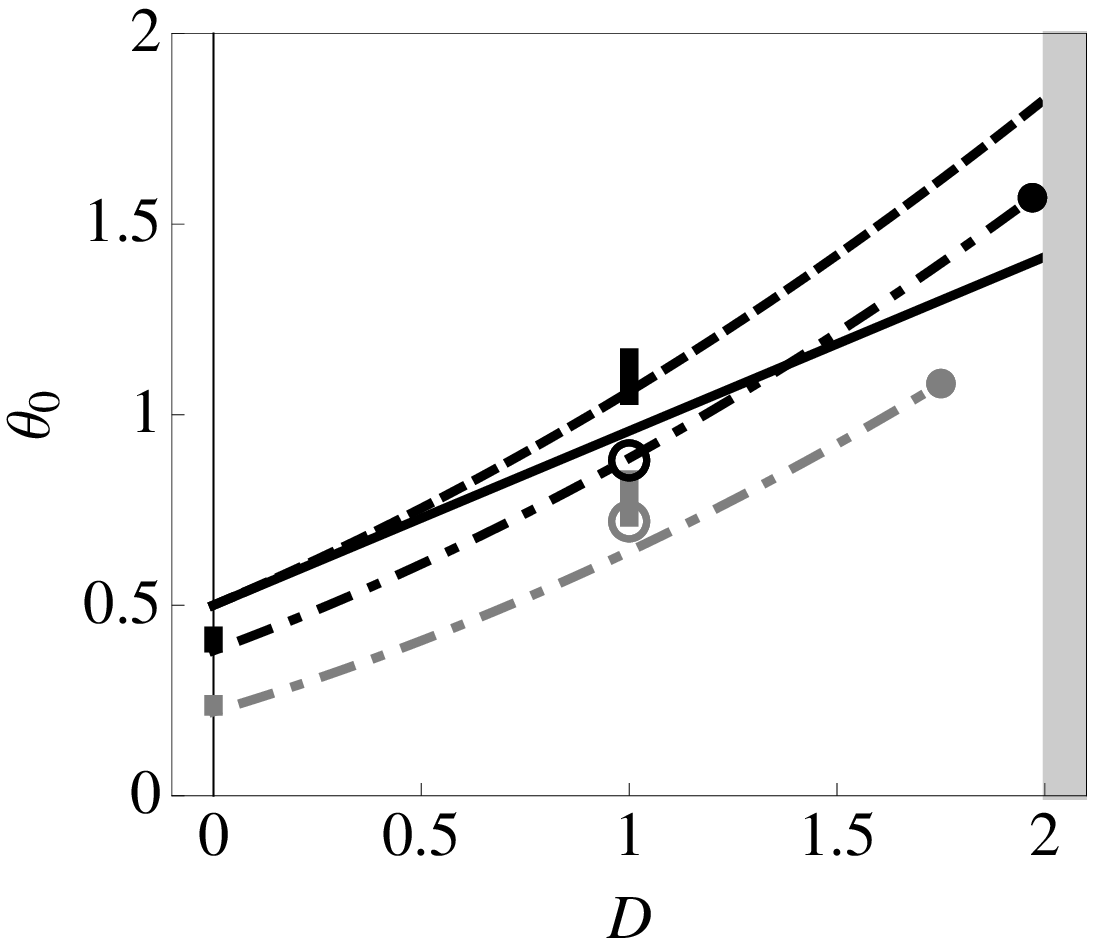}
\\[2mm]
(a)&& (b)
\end{tabular}
\caption{Persistence exponents  $\theta_{\infty,0}(d,d-D)$ of  the global order parameter of a manifold of codimension $D$ as a function of $D$, in the case of (a) non-vanishing and (b) vanishing  initial value of the order parameter. 
The dash-dotted lines in (a) and (b) correspond to the analytical estimates of $\theta_{\infty,0}$ which  account perturbatively for the effects of non-Gaussian fluctuations  (c.f., Sec~\protect{\ref{sec:BGaux}}) within the Ising universality class with Glauber dynamics in $d=3$ (black) and $d=2$ (grey). These lines terminate on the right at $D=2-\eta$, as the asymptotic decay of the persistence probability of this model is no longer algebraic for a codimension larger than these values. 
Apart from the addition of these dash-dotted lines, this figure is the same as Fig.~\protect{\ref{fig:Gaux}}. 
\label{fig:beyondGaux}}
\end{figure}

Here we focus on the $O(n=1)$, Ising universality class and we consider both the cases $m_0=0$ and $m_0\neq 0$, for which the constants $A$ and $B$ in Eq.~\reff{eq:RepsD}  actually take different values.  Comparing the expression \reff{eq:RepsD} of ${\mathcal R}$ for $\eps=0$ with Eqs.~\reff{eq:est2} and~\reff{eq:est2_0} one readily finds that
\begin{equation}
A = 
\begin{cases}
\sqrt{2}-1 = 0.414\ldots & \mbox{for}\quad m_0 = 0\,,\\
\phantom{\sqrt{2}-1 = } \; 0.723\ldots & \mbox{for}\quad m_0 \neq 0\,.
\end{cases}
\end{equation}
Analogously, the coefficient $B$ for the cases  $m_0=0$ and $m_0\neq 0$ can be determined by comparison with the results for ${\mathcal R}(4-\eps,D=0)$ presented in Eq.~(12) of Ref.~\cite{us_epl} and in Eq.~(19) of Ref.~\cite{oerding_persist}, respectively:
\begin{equation}
B = 
\begin{cases}
0.00754\ldots & \mbox{for}\quad m_0 = 0\,,\\
0.0273\ldots &  \mbox{for}\quad m_0 \neq 0\,.
\end{cases}
\end{equation}
(Note, however, that for $m_0=0$ the first term of the expansion of  ${\mathcal R}(4-\eps,D=0)-1$ is of order $\eps^2$~\cite{oerding_persist}.) These values result in
\begin{equation}
{\mathcal R(4-\eps,D)} = 
\begin{cases}
{\mathcal R}_0\;\,\equiv  1+  D\times 0.414\ldots+ \eps^2\times 0.00754\ldots+ {\cal O}(D^2,\eps^2 D,\eps^3)&  \mbox{for}\quad m_0 = 0\,,\\
{\mathcal R}_\infty\equiv 1 + D\times  0.723\ldots + \phantom{^2}\eps\times 0.0273\ldots\phantom{4}+ {\cal O}(D^2,\eps D,\eps^2) & \mbox{for}\quad  m_0 \neq 0\,,
\end{cases}
\label{eq:Reps}
\end{equation} 
which clearly show that the corrections due to a finite codimension are quantitatively more relevant than those due to non-Gaussian fluctuations. In particular the dependence of the latter on the dimensionality $d=4-\eps$ is weak enough that a simple extrapolation of Eq.~\reff{eq:Reps} to $\eps=2$ and $\eps=1$  should provide quantitatively reliable estimates of ${\mathcal R}$ as a function of $D$ for $d=2$ and $d=3$, respectively.  These estimates, denoted by the superscript \bg,  are reported in Tab.~\ref{tab:exp}.
The Markovian approximation $\mu(d,D)$ of the persistence exponents in the two cases $m_0\neq 0$ and $m_0=0$, is respectively given by Eqs.~\reff{eq:hyp_b} and~\reff{eq:hyp0}, where the critical exponents $\eta$, $z$, and $\theta_{\rm is}$ in spatial dimension $d=2$ and $3$ (see, \eg, Refs.~\cite{PV,critical_review}) take the values which are reported in Tab.~\ref{tab:exp} together with the resulting expressions for  $\mu_{0,\infty}$ as a function of $D$. 
\begin{table}
\begin{tabular}{||c||c|c||}
\hline
\hline
&$d=2$&$d=3$ \\
\hline
\hline
${\mathcal R}_0^{\bg}$ & $1.03 + 0.414 D$ & $1.01 + 0.414 D$\\
${\mathcal R}_\infty^{\bg}$ & $1.05 + 0.723 D$ & $1.03 + 0.723 D$\\
\hline
$\eta$ & 1/4 & 0.03\\
$z$ & 2.17 & 2.04\\
$\theta_{\rm is}$ & 0.38 & 0.14\\
\hline
$\mu_0$ & $0.216 + 0.23 D$ & $0.377 + 0.245 D$\\
$\mu_\infty$ & $1.46 \,\,\,+ 0.23 D$ & $1.74\,\,\,+ 0.245 D$\\
\hline
\hline
\end{tabular}
\caption{Estimates of the universal ratios ${\mathcal R}_{0,\infty}^\bg$ [up to ${\cal O}(D^2)$, see Eq.~\reff{eq:Reps}], of the critical exponents (see, \eg, Refs.~\cite{PV,critical_review}), and of the resulting Markovian approximations $\mu_{0,\infty}$ [Eqs.~\reff{eq:hyp0} and~\reff{eq:hyp_b}] for the persistence exponents $\theta_{0,\infty}$ within the Ising universality class  in two and three spatial dimensions. The analytical estimates $\theta^\bg_{0,\infty} \equiv {\mathcal R}^\bg_{0,\infty} \times \mu_{0,\infty}$ of $\theta_{0,\infty}$ agree with those of Refs.~\cite{oerding_persist,us_epl} for $D=0$, whereas for $D=1$ they yield the values reported in, c.f., Tab.~\protect{\ref{tab:expMC}}.
}
\label{tab:exp}
\end{table}
The persistence exponents $\theta_0$ and $\theta_\infty$  can 
be estimated as $\theta_0^\bg \equiv {\mathcal R}_0^\bg\times \mu_0$ and $\theta_\infty^\bg \equiv {\mathcal R}^\bg_\infty\times \mu_\infty$, which are reported as dash-dotted lines in the two panels of Fig.~\ref{fig:beyondGaux} for $d=2$ (gray) and $d=3$ (black) together with the Monte Carlo and Gaussian estimates anticipated in the corresponding Fig.~\ref{fig:Gaux}. The comparison among all these estimates will be presented further below in Sec.~\ref{sec:MCth}.

%
%
%
%
%

%
In order to improve on the analytical estimates of the persistence exponents and to go beyond the linear dependence on $d$ and $D$ expressed by Eq.~\reff{eq:Reps} (see also Tab.~\ref{tab:exp}), one would need first of all an expression of ${\mathcal R}$ which accounts for non-Markovian corrections beyond the leading order, \ie, an extension to higher orders of the perturbation theory developed in Refs.~\cite{majumdar_critical,oerding_persist}. Secondly, an analytic expression for the contribution of non-Gaussian fluctuations to  $C_\bQ(t,t')$ would be required.
In the present work, instead of pursuing this strategy, we shall investigate the general features of the $D$-dependence of the persistence exponent $\theta_\infty$ on the basis of the Monte Carlo results presented in the next section and of the analytical study of  transverse fluctuations in the $O(n\rightarrow\infty)$, presented in Sec.~\ref{sec:On}.

\section{Monte Carlo results}

\label{sec:MC}

In order to test our prediction of a temporal crossover in the critical persistence of manifolds for $\zeta < 0$  (see Tab.~\ref{fig_table}) as well as the  theoretical estimate of $\theta_\infty$ as a function of $d$ for $D=1$ 
--- see Eq.~\reff{theta_smallD} and Tab.~\ref{tab:exp} --- we studied   via Monte Carlo simulations the ferromagnetic
Ising model with spins $s_i = \pm 1$ and Glauber dynamics in two and three spatial dimensions. 

\subsection{Line magnetization within the two-dimensional Ising-Glauber model}
\label{sec:MC2d}
\begin{figure}
\begin{tabular}{ccc}
\begin{minipage}{0.4\linewidth}
 \includegraphics[angle=-90,width=\linewidth]{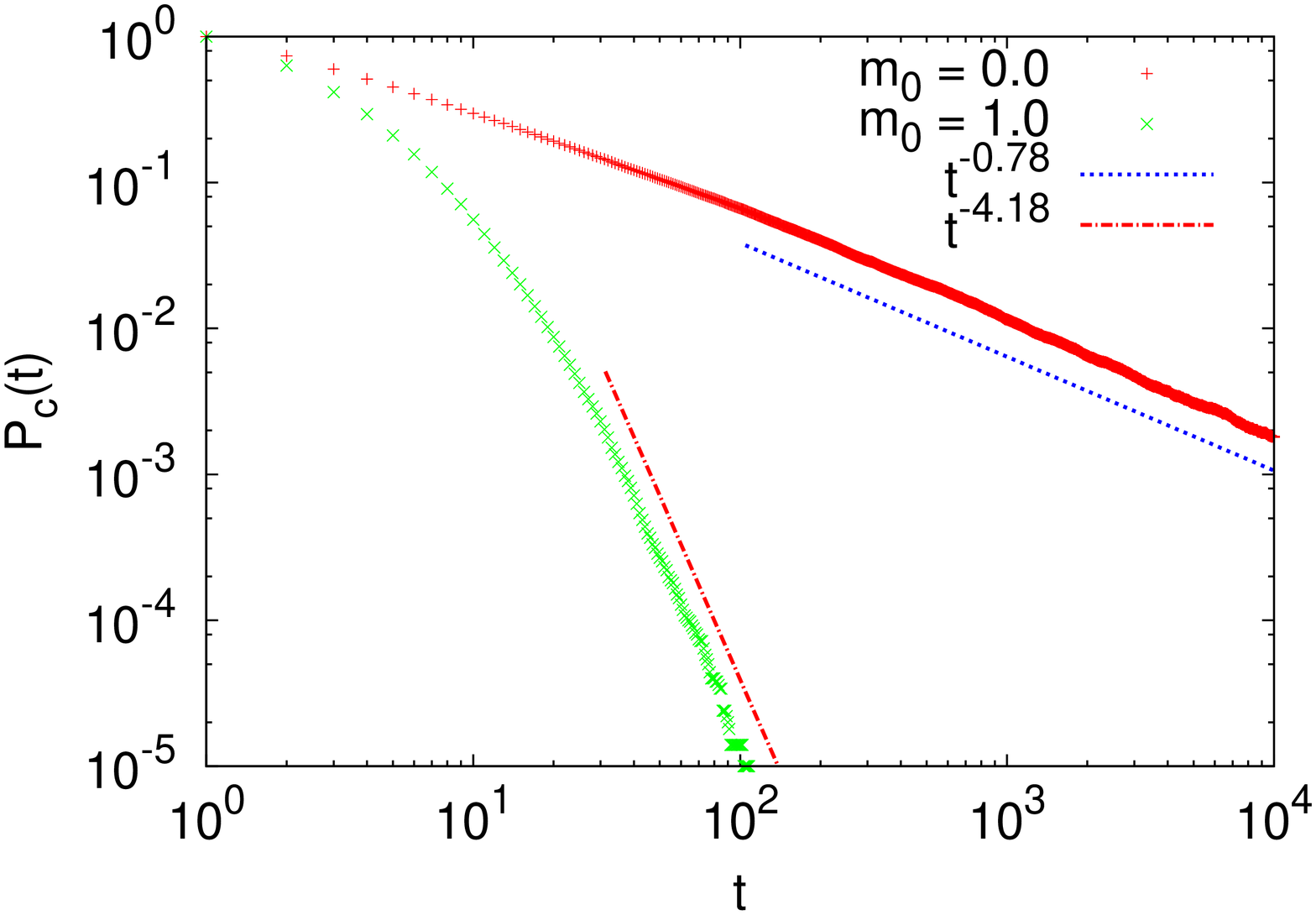}
\end{minipage}
&\quad&
\begin{minipage}{0.4\linewidth}
 \includegraphics[angle=-90,width = \linewidth]{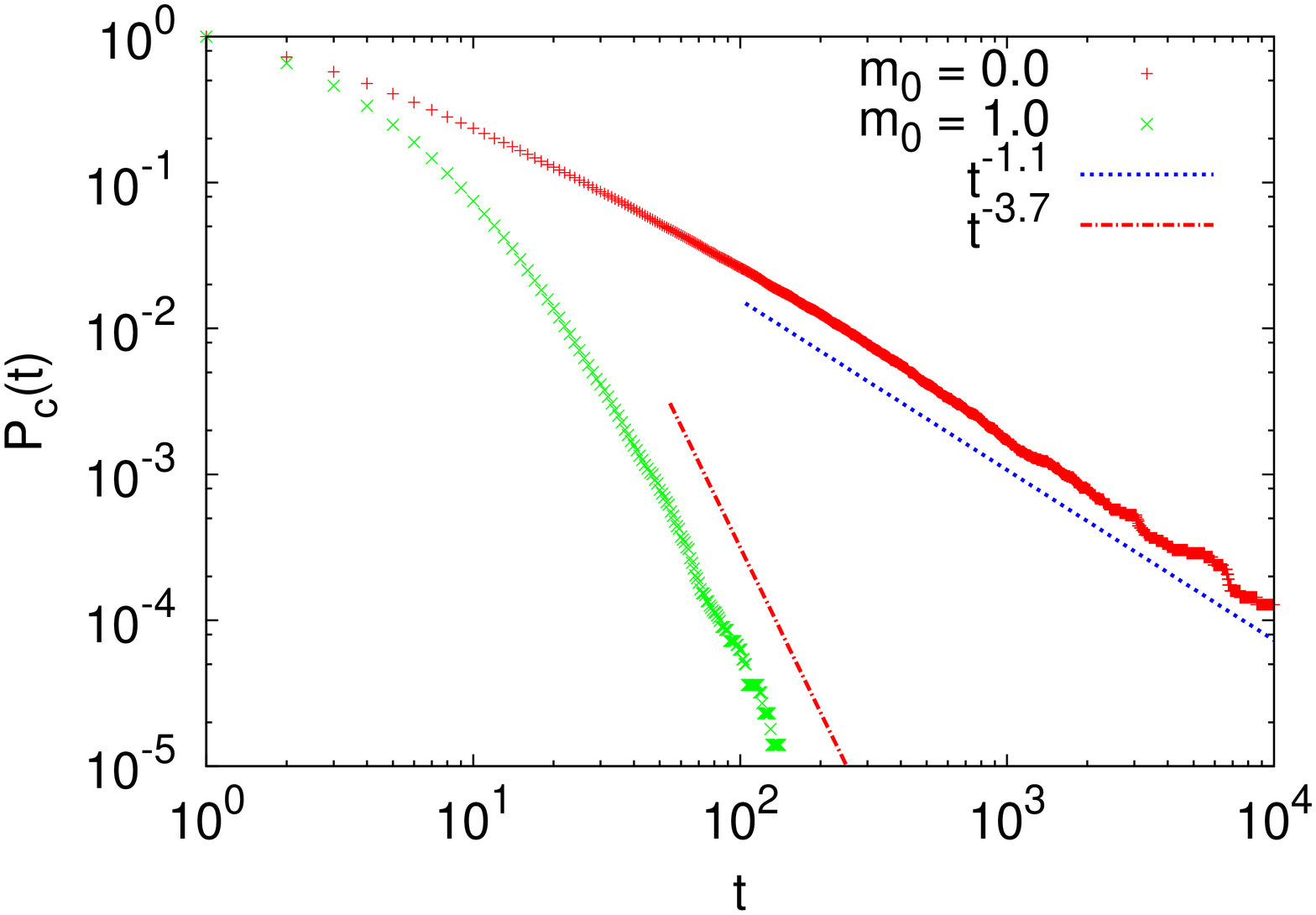}
\end{minipage}\\
(a) &\quad& (b)
\end{tabular}
\caption{
Persistence probability $P_c(t)$ of the magnetization of (a) a line and (b) a plane within, respectively, the two- and three-dimensional  critical Ising model with Glauber dynamics for the two extreme values $m_0=0$ (top data set) and $m_0=1$ (bottom data set) of the initial magnetization $m_0$. These data sets correspond, respectively, to the two different regimes $t \ll \tau_m$ and $t \gg \tau_m$, within which  $P_c(t)$ decays algebraically at large times but with  distinct exponents $\theta_0^{\mc}$ (top) and $\theta_\infty^{\mc}$ (bottom), the best fit values of which are indicated by the associated straight lines.
These numerical data provide clear evidence for the occurrence of the dynamical crossover discussed in Sec.~\ref{sec:model_scal}. 
%
} 
\label{fig_crossover_01}
\end{figure}

We first focus on the two-dimensional case of a $L\times L$ square lattice with periodic boundary conditions, for which the size $L$ ranges from $64$ to $256$ and it is such that the data presented below are not appreciably affected by the expected finite-size corrections. We study the temporal evolution of the fluctuating magnetization of a complete line (row or column, \ie, $d'=1$) selected within the lattice and we calculate the associate persistence probability. 
Taking into account the values of the exponents $z$ and $\eta$ reported in Tab.~\ref{tab:exp}, this choice corresponds to $\zeta = (d-d'-2+\eta)/z \simeq -0.345 < 0$ and therefore the persistence probability of the line magnetization is expected to decay algebraically at large times. 
The system is initially prepared in a random configuration with $N_{+}$ up and $N_{-}$ down spins, where $N_{\pm} = L^2 (1 \pm m_0)/2$. Then, at each subsequent time step, a site is randomly chosen and the move $s_i \mapsto - s_i$ is accepted or rejected according to Metropolis rates corresponding to the critical temperature $T=T_c$. One time unit corresponds to $L^2$ attempted updates of spins.
The determination of the persistence probability $P_{c}(t)$ of the fluctuations requires also the
knowledge of the global magnetization $M(t) = L^{-2} \langle \sum_i s_i\rangle$, 
which we obtained by averaging over 2000 realizations of the dynamics. For each of these realizations, we also choose a new random initial condition with fixed magnetization $m_0$. The persistence probability is then computed as the
probability that the fluctuating magnetization of the line $L^{-1}\sum_{i\in{\rm line}} s_i - M(t)$ has not changed sign between $t=0$ and time $t$. This probability is determined on the basis of more than
$10^6$ samples. In Fig.~\ref{fig_crossover_01}(a), we show the results of our simulations corresponding to $L=256$ and to two different values of the magnetization $m_0$. The top curve refers to $m_0 = 0$, \ie, $\tau_m \to \infty$, such that the system is always in the regime $t \ll \tau_m$  investigated in Ref.~\cite{manifold}. These data are fully compatible with a power-law decay with an exponent $\theta_0^\mc(d=2,d'=1) = 0.78(5)$ (see also Tab.~\ref{tab:expMC}), which turns out to be in a rather good agreement with the value $\simeq 0.72$ reported in Ref.~\cite{manifold}. 
The bottom curve in Fig.~\ref{fig_crossover_01}(a), instead, corresponds to $m_0 = 1$. In this case, $\tau_m \ll 1$ and after a short initial transient, the system enters the regime $t \gg \tau_m$, within which the persistence probability is expected, according to Sec.~\ref{sec:model_scal}, to decay algebraically at large times with an exponent $\theta_\infty$. The numerical data in the figure are compatible with such a power-law decay, characterized by an exponent $\theta_\infty^\mc(d=2,d'=1) \simeq 4.2(1)$ which is significantly larger than the one measured in the case $m_0=0$. 
%
%
%
%
%
%

\subsection{Plane magnetization within the three-dimensional Ising-Glauber model}
\label{sec:MC3d}

In order to investigate the dependence of the persistence exponent on the space dimensionality $d$ at fixed codimension $D$, we extended the investigation presented above ($d=2$, $D=1$) to the three-dimensional case, focusing on the magnetization of a plane ($d=3$, $D=1$) within the Ising-Glauber model on a $L\times L \times L$ cubic lattice with periodic boundary conditions. The size $L$ ranges from $16$ to $128$ and it is such that the data presented below are not appreciably affected by the expected finite-size corrections. 
Taking into account the values of the exponents $z$ and $\eta$ 
reported in Tab.~\ref{tab:exp}, this choice corresponds to $\zeta = (D-2+\eta)/z \simeq -0.48 < 0$
and therefore the persistence probability is expected to decay algebraically at large times. 
In Fig.~\ref{fig_crossover_01}(b), we report the results of our
simulations on a lattice with $L=128$, for the two extreme values $m_0=0$ (top data set) and $m_0=1$ (bottom data set) of the initial magnetization $m_0$, as we did in panel (a) of the same figure for the case of the magnetization of a line within the two-dimensional model. At large times one clearly observes that the power of the algebraic decay changes  when passing from $m_0=0$ to $m_0=1$. In the former case (top curve)  $t \ll \tau_m$ and the data
are compatible with $P_c(t) \sim t^{-\theta_0(d=3,d'=2)}$ with an exponent $\theta_0^{\mc}(d=3,d'=2) \simeq 1.10(5)$. This value is significantly larger than the estimate $\theta_0^{\mc}(d=3,d'=2) \simeq 0.88$ preliminarily reported in Ref.~\cite{manifold}. However, such an estimate was obtained for rather small system sizes $L = 15$ and $31$ and therefore it might be biased by finite-size effects.
The bottom curve in Fig.~\ref{fig_crossover_01}(b) corresponds  to the case
$t \gg \tau_m$  and the numerical data for the persistence probability still decay  algebraically at large times, but 
with an exponent $\theta^{\mc}_\infty(d=3,d'=2) \simeq 3.7(1)$ which is significantly larger than in the case $m_0=0$. 

\begin{figure}[h]
\begin{tabular}{ccc}
\begin{minipage}{0.45\linewidth}
 \includegraphics[angle=-90,width = \linewidth]{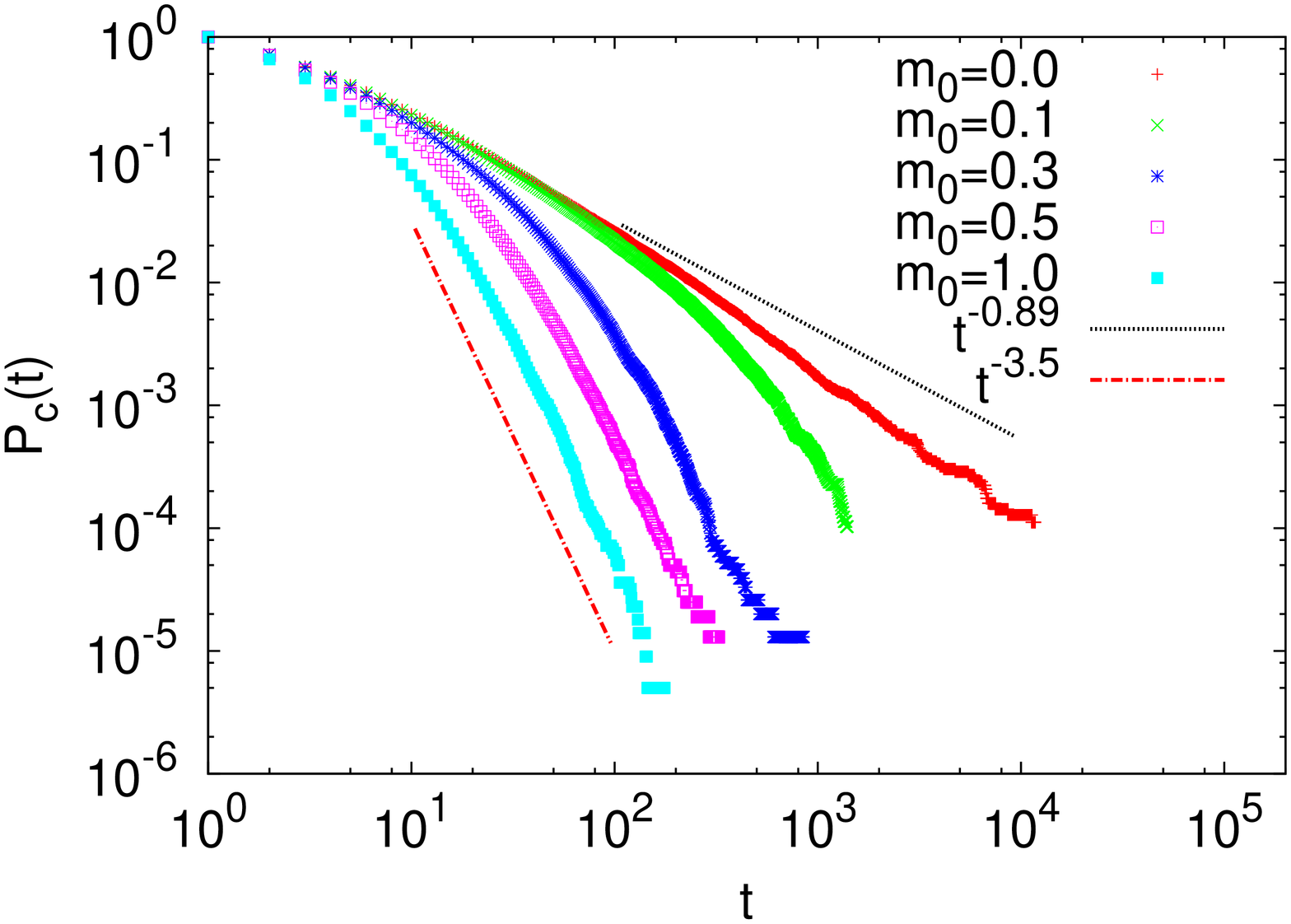}
\end{minipage}
&\quad&
\begin{minipage}{0.45\linewidth}
 \includegraphics[angle=-90,width=\linewidth]{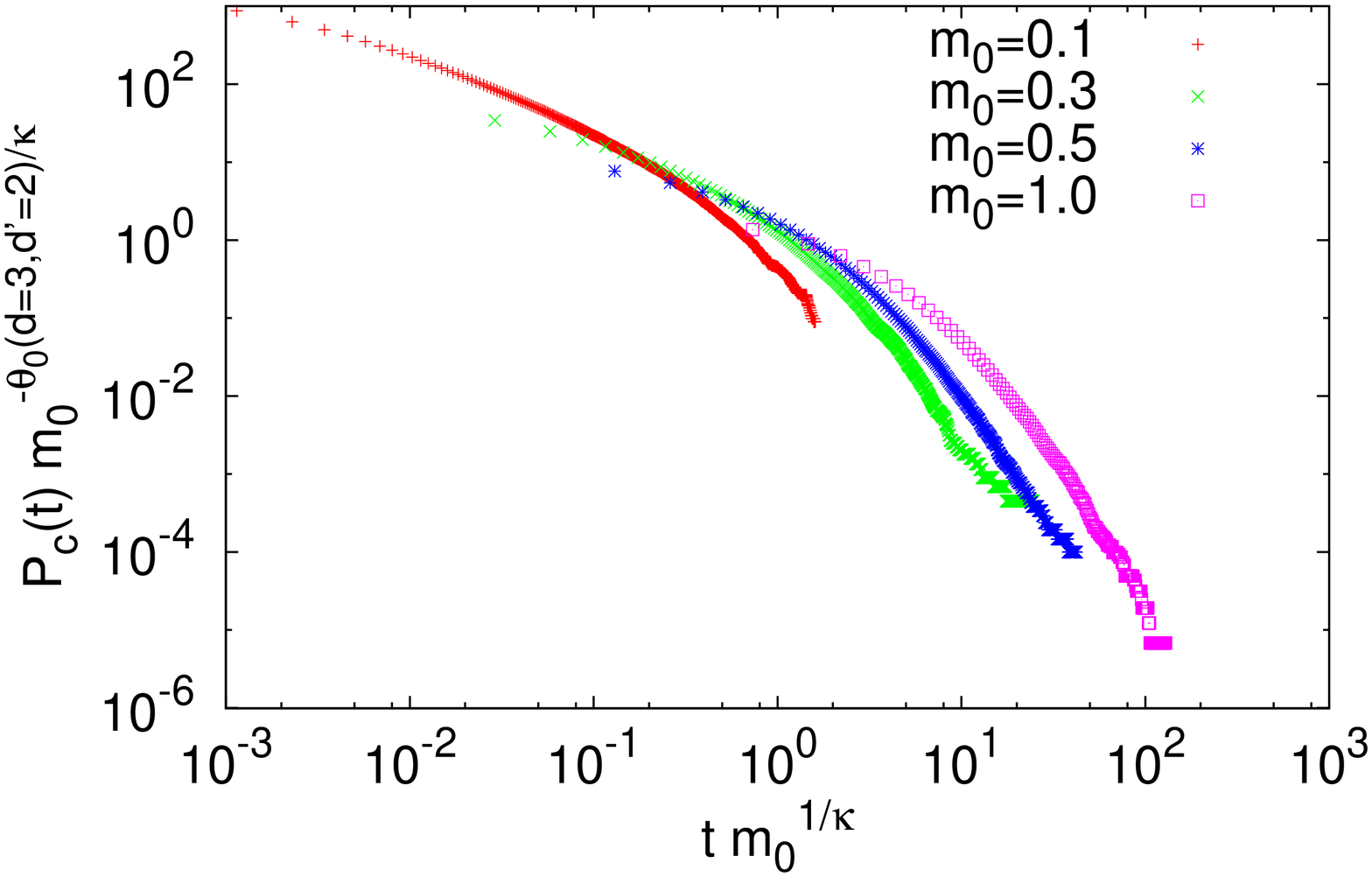}
\end{minipage}
\\
(a) && (b)
\end{tabular}
\caption{Critical persistence probability $P_c(t)$ of the fluctuating
  magnetization of a plane within the three-dimensional Ising model for different
  values of the initial magnetization $m_0$. In panel (a) $P_c$ is plotted as a function of time $t$ and its  long-time algebraic decay clearly shows a crossover, which is controlled by the initial value $m_0$ of the magnetization. The uppermost and lowermost data sets are the same as those presented in Fig.~\protect{\ref{fig_crossover_01}}(b), whereas the associated straight lines indicate algebraic decays with the exponents $\theta^\bg_0$ and $\theta^\bg_\infty$, respectively, analytically estimated in Sec.~\protect{\ref{sec:BGaux}} (see Fig.~\protect{\ref{fig:beyondGaux}} and, c.f., Tab.~\protect{\ref{tab:expMC}}).
In panel (b) the data presented in (a) are scaled according to the heuristic scaling ansatz discussed in the main text: 
The data for $P_c(t)\times m_0^{-\theta_0(d=3,d'=2)/\kappa }$ should collapse onto a single master curve when plotted as a function of $t\times m_0^{1/\kappa}$ with $\theta_0(d=3,d'=2) = \theta_0^{\mc}(d=3,d'=2) = 1.1$ and $\kappa = 0.34$ [see Eq.~\reff{eq:kappa}]. However, this is clearly not the case and a significant dependence on $m_0$ is still observed in the resulting scaled curves.}
\label{fig_nocollapse}
\end{figure}

\begin{figure}[h]
 \includegraphics[angle=-90,width=0.5\linewidth]{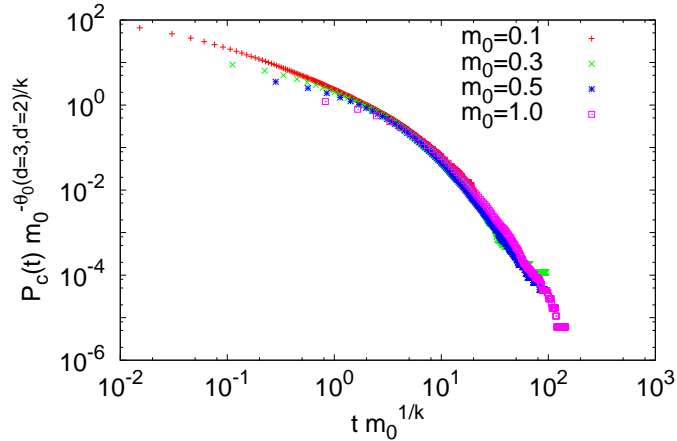}
\caption{Effective scaling of the Monte Carlo data presented in Fig.~\ref{fig_nocollapse}(a) for the critical persistence probability $P_c(t)$ of the fluctuating magnetization of a plane within the three-dimensional Ising model. Data corresponding to different values of the initial magnetization $m_0$ collapse onto a single master curve when 
$P_c(t) \times m_0^{-\theta_0(d=3,d'=2)/k}$ is plotted as a function of $t m_0^{1/k}$ with $\theta_0(d=3,d'=2) = 1.1$ 
and $k = 0.54 \neq \kappa$.}
\label{fig_master_curve}
\end{figure}

Intermediate values $0 < m_0 < 1$ of $m_0$, correspond to finite and non-vanishing $\tau_m$ and therefore
one expects $P_c(t)$ to display the two different power-law behaviors described separately above within the two consecutive time ranges $t\ll \tau_m$ and $t\gg \tau_m$ . Indeed, this  is clearly displayed in Fig.~\ref{fig_nocollapse}(a) where we report the time dependence of
the persistence probability $P_c(t)$ of the magnetization of a plane in three dimensions, for various values of
$m_0$: At relatively small times (\ie, larger than some microscopic scale but smaller than $\tau_m$) $P_c(t)$ decreases algebraically with the power $\theta_0$ characteristic of the decay of the curve corresponding to $m_0=0$. As time $t$ increases and exceeds $\tau_m$, however, one observes a crossover towards an algebraic decay with the power  $\theta_\infty$ characteristic of the decay of the persistence probability for $m_0=1$.
(For comparison, in Fig.~\ref{fig_nocollapse}(a) we also indicate the straight lines corresponding to algebraic decays with the powers $\theta_{0,\infty}=\theta^\bg_{0,\infty}$ predicted in Sec.~\protect{\ref{sec:BGaux}} and reported in Fig.~\protect{\ref{fig:beyondGaux}} and, c.f., Tab.~\protect{\ref{tab:expMC}}.)
As the effect of a finite $m_0$ is to introduce a time scale $\tau_m$ into the problem,
it is natural to wonder whether these curves corresponding to different
values of $m_0$ are characterized by dynamical scaling, \ie, if they collapse onto
a single master curve after a proper rescaling of the time $t$ and of the probability $P_c(t)$ involving $\tau_m$. 
The natural heuristic candidate for such a scaling form is
$P_c(t) \sim \tau_m^{-\theta_0} {\mathcal P}(t/\tau_m)$, where, for consistency with the known behaviors for 
$t \ll \tau_m$ and $t \gg \tau_m$, one has ${\mathcal P}(x\ll 1)\sim x^{-\theta_0}$ and ${\mathcal P}(x\gg 1)\sim x^{-\theta_\infty}$.
This scaling ansatz can be tested by plotting $\tau_m^{\theta_0} P_c(t)$ or, equivalently, 
$m_0^{-\theta_0/\kappa}P_c(t)$, as a function of $t/\tau_m$ or, equivalently,
$m_0^{1/\kappa}t$ where the exponents $\theta_0=\theta_0(d,d')$ and $\kappa(d)$ are the appropriate ones to the dimensionality of the model and of the manifold.  
In Ref.~\cite{us_epl} it was shown that the persistence probability of the global
magnetization ($d'=d$) of the two-dimensional Ising-Glauber model obeys indeed such a scaling form with
the expected value of $\kappa = \theta'_{\rm is} + \beta/(\nu z) \simeq 0.249$. This numerical value follows form Eq.~\reff{eq:kappa},  where one uses the values of the exponents reported in Tab.~\ref{tab:exp} together with the fact that in two spatial dimensions $\beta/\nu =\eta/2$. 
Similarly, we have checked (data not shown) that the same scaling behavior holds 
for the global magnetization of the three-dimensional Ising model, with the expected value of 
$\kappa = 0.34(1)$, which follows from Eq.~\reff{eq:kappa} with $\theta'_{\rm is}= 0.104(3)$ \cite{Gr-95}, $\beta = 0.3267(10)$, $\nu = 0.6301(8)$ \cite{blote}, $z = 2.04(2)$ \cite{Gr-95}. 
However, the present numerical data indicate that this heuristic
scaling ansatz does not capture the actual behavior of the persistence probability for the fluctuating magnetization of a
manifold, neither in the case of a line in two dimensions nor of a plane in three dimensions. 
The data for a plane in three dimensions are shown in Fig.~\ref{fig_nocollapse}(a), where we plot $m_0^{-\theta_0(d=3,d'=2)/\kappa}P_c(t) $ as a function of $m_0^{1/\kappa}t$ with $\theta_0(d=3,d'=2) = \theta^{\mc}_0(d=3,d'=2) =1.1$ [see Fig.~\ref{fig_crossover_01}(b)] and $\kappa = 0.34$ as derived above:
clearly the curves corresponding to different values of the magnetization $m_0$ do not collapse onto a single master curve. We have carefully checked that the absence of data collapse is not due to finite-size effects. 
%
%
%
%
%
%
%
This lack of scaling can be qualitatively understood on the basis of the fact that the persistence probability we have discussed so far is, in fact, a special case of a {\it two-time} quantity. 
Indeed, consider the persistence probability $\bar P_c(t,t')$, defined as the probability that
the process $X(\br, t)$ does not change sign between the times $t'$ and
$t > t'$. In terms of this quantity, the persistence probability $P_c(t)$ studied above is given by $P_c(t) = \bar P_c(t,\tmic)$, where $\tmic$ is some non-universal microscopic time scale set, \eg, by the elementary moves of the dynamics. 
The scaling form of the correlation function $\langle X(\br, t) X(\br, t')
\rangle \equiv {\cal C}(t/\tau_m, t'/\tau_m)$ [see Eq.~\reff{expr_gauss_complete}]
implies straightforwardly 
that $\bar P_c(t,t') \equiv {\bar{\cal P}}(t/\tau_m,t'/\tau_m)$, where $\tau_m \propto m_0^{-1/\kappa}$.
In addition, the scaling behavior of the correlation function becomes independent of $\tau_m$ within the 
following two different regimes --- which can be investigated
analytically by taking, respectively, the limits $\tau_m\rightarrow\infty$ and $\tau_m\rightarrow 0$:  
(I) $t'<t \ll \tau_m$ within which $\langle X(\br, t) X(\br, t')
\rangle = {\cal C}^{\rm I}(t/t')$ and (II)  $\tau_m \ll  t'<t$ within which
$\langle X(\br, t) X(\br, t') \rangle = {\cal C}^{\rm II}(t/t')$. 
Correspondingly, the persistence probability ${\bar P}_c(t,t')$  takes two different scaling forms $\bar{\cal P}$:
\begin{eqnarray}
{\rm (I)} && \quad {\bar P}_c(t,t') \equiv {\bar{\cal P}}^{\rm I}(t/t')
  \stackrel{t\gg t'}{\sim} (t/t')^{-\theta_0(d,d')} \quad\mbox{for}\quad  t' < t \ll \tau_m, \label{regime_I} \\
{\rm (II)}  &&\quad {\bar P}_c(t,t') \equiv {\bar{\cal
  P}}^{\rm II}(t/t') 
  \stackrel{t\gg t'}{\sim} (t/t')^{-\theta_\infty(d,d')} \quad\mbox{for}\quad \tau_m \ll  t'<t.  \label{regime_II} 
\end{eqnarray}
In passing we mention that, interestingly enough, two analogous regimes emerge in the study of
the persistence of fluctuating interfaces, in which the role of $\tau_m$ is played by the equilibration time $\tau_{eq} \sim L^2$ for a system of finite size $L$ \cite{krug_ew}.
As anticipated above, the case $m_0 = 0$ corresponds to regime (I) [Eq.~\reff{regime_I}], whereas  $m_0 = 1$ to regime (II) [Eq.~\reff{regime_II}]. While these two instances are well understood, our simulations for
intermediate values of $m_0$ are likely to correspond to a {\it third}
regime $t' \ll \tau_m\ll t$ which we are presently unable to
describe analytically for generic $D > 0$ because the correlation function of the process cannot be made stationary due to its residual dependence on $\tau_m$. [Note that the introduction of the logarithmic time renders correlation functions stationary as long as they depend only on $t'/t$ or, more generally, on ratios $f(t')/f(t)$, see further below.]
In particular there is no obvious reason
why the scaling  $P_c(t) \sim t^{-\theta_0} {\mathcal P}(t/\tau_m)$ or, equivalently,
$P_c(t) \sim  m_0^{\theta_0(d,d')/\kappa} P_c(m_0^{1/\kappa} t)$
--- which well describes the behavior of $P_c(t)$ for $D=0$~\cite{us_epl} --- 
should actually hold in general and therefore one should not be surprised by the fact
that this scaling form turns out to be violated in the case of Fig.~\ref{fig_nocollapse}(b). 
Within the Gaussian approximation, however, one can heuristically understand why the case $D=0$ investigated in Ref.~\cite{us_epl} is indeed special in this respect. In fact, from Eq.~\reff{expr_gauss_complete} one can see that only for $D=0$ the correlation function  $\langle X(\br, t) X(\br, t') \rangle$ has the form $\langle X(\br, t) X(\br, t')
\rangle = L(t'/\tau_m)/L(t/\tau_m)$, with $L(\tilde t) = {\tilde t}^{\,1/2} \sqrt{I_0(...,\tilde t)}$, such that $L(\tilde t)\sim {\tilde t}^{\,1/2}$ for  $\tilde t \ll 1$, whereas  $L(\tilde t)\sim {\tilde t}^{\,2}$ for  $\tilde t \gg 1$ --- we do not specify here the first argument of $I_0$ as $I_D(x,u)$ is actually independent of $x$ for $D=0$, see Eq.~\reff{expr_gauss_complete_b}. 
The persistence probability for such a process can be calculated exactly~\cite{slepian}, 
after a mapping onto a stationary process: 
$\bar P_c(t,t') = (2/\pi) {\rm arcsin}[L(t'/\tau_m)/L(t/\tau_m)]$.
Therefore in the third regime $t' \ll \tau_m \lesssim t$ one has 
$\bar P_c(t,t') \sim (t'/\tau_m)^{1/2}/L(t/\tau_m)$
which actually renders $P_c(t) = {\bar P}_c(t,t'=\tmic)
\sim  \tau_m^{-\theta_0} {\cal P}(t/\tau_m)$, \ie, the heuristic scaling form anticipated above, with the correct exponents $\theta_0=1/2$ and $\theta_\infty=2$~\cite{us_epl}. Finally, even 
though such a scaling form does not work for $D > 0$ our numerical
simulations indicate that data corresponding to different values of $m_0$ actually 
fall --- to a rather good extent, see Fig.~\ref{fig_master_curve} --- onto a single master curve obtained by plotting 
$m_0^{-\theta_0(d=3,d'=2)/k}P_c(t)$ as a function of 
$m_0^{1/k} t$ with $\theta_0(d=3,d'=2) = 1.1$
and $k = 0.54 \neq \kappa$. However, the origin of this effective scaling is presently unclear and it surely deserves 
further investigations. 
%
%
%
%
%

\subsection{Comparison between analytical and numerical results}
\label{sec:MCth}

The comparison between the analytical predictions of Sec.~\ref{sec:BGaux} and the available numerical estimates of $\theta_{0,\infty} $ discussed in Secs.~\ref{sec:MC2d} and~\ref{sec:MC3d} or reported in the literature is summarized in Fig.~\ref{fig:beyondGaux} and, for $D=1$, also in Tab.~\ref{tab:expMC}.
In particular, panels (a) and (b) of Fig.~\ref{fig:beyondGaux} compare the various estimates of $\theta_\infty$ and $\theta_0$, respectively, as  functions of the codimension $D$, in spatial dimension $d=2$ (grey lines and markers) and $d=3$ (black lines and markers).
As already pointed out in Refs.~\cite{us_epl,oerding_persist}, for vanishing codimension $D=0$ the agreement between the Monte Carlo estimate $\theta_{\infty,0}^\mc$ (vertical bars in Fig.~\ref{fig:beyondGaux}) and the analytical estimates  
$\theta_{\infty,0}^\bg$ (dash-dotted lines)  of $\theta_{\infty,0}$  is very good both in $d=2$ and $d=3$. 
For $D=1$, instead, the comparison is less clear. On the one hand, our Monte Carlo estimate of $\theta_\infty^{\mc}(d=3,d'=2)=3.7(1)$ in $d=3$ (indicated in Fig.~\ref{fig:beyondGaux}(a) as a vertical black bar --- see Sec.~\ref{sec:MC3d} for details and Tab.~\ref{tab:expMC}), is in a rather good agreement with the corresponding analytical estimate $\theta_\infty^\bg(d=3,d'=2)=3.5$ reported in Tab.~\ref{tab:expMC} [dash-dotted black line in Fig.~\ref{fig:beyondGaux}(a)] and also with the result of the Gaussian approximation (solid and dashed lines). On the other hand, $\theta_\infty^\mc(d=2,d'=1)=4.2(1)$ in $d=2$  (vertical gray bar in Fig.~\ref{fig:beyondGaux}(a); see Sec.~\ref{sec:MC2d} for details) is significantly larger than the corresponding analytical estimate $\theta^\bg_\infty(d=2,d'=1)=3.0$  reported in Tab.~\ref{tab:expMC} [dash-dotted gray line  in Fig.~\ref{fig:beyondGaux}(a)]. In addition, the fact that $\theta_\infty^{\mc}(d=2,d'=1)>\theta_\infty^{\mc}(d=3,d'=2)$ is at odd with what is observed for all the other numerical and analytical estimates presented in this work (\cf, Fig.~\ref{fig:On} for the $O(n\rightarrow\infty)$ universality class), \ie, that $\theta_{\infty,0}(d=2, d'=1)<\theta_{\infty,0}(d=3,d'=2)$. 
\begin{table}
\begin{tabular}{||l||l|l||}
\hline
\hline
&\multicolumn{2}{|c||}{$D=1$}\\
\hline
\hline
&\multicolumn{1}{|c|}{$d=2$} & \multicolumn{1}{|c||}{$d=3$} \\
\hline
\hline
$\theta_0^\mc$ & $0.78(5)$ & $1.10(5)$ \\
$\theta_0^\bg$ & $0.64$ & $0.89$ \\
\hline
$\theta_\infty^\mc$ & $4.2(1)$   & $3.7(1)$\\
$\theta_\infty^\bg$  & $3.0$   & $3.5$ \\
\hline
\hline
\end{tabular}
\caption{Monte Carlo and analytical estimates for the persistence exponents $\theta_0$ and $\theta_\infty$ of the magnetization of a manifold of codimension $D=1$ --- \ie, a line in spatial dimension $d=2$ or a plane in $d=3$ --- for an Ising model with Glauber dynamics, relaxing from an initial state with $m_0=0$ and $m_0\neq 0$, respectively.  The corresponding Monte Carlo estimates $\theta_{0,\infty}^\mc$ have been obtained as described in Sec.~\protect{\ref{sec:MC}}, whereas the analytical estimates $\theta^\bg_{0,\infty} \equiv {\mathcal R}^\bg_{0,\infty} \times \mu_{0,\infty}$ follow from the direct extrapolation to $D=1$ of the results of Sec.~\protect{\ref{sec:BGaux}}, summarized in 
Tab.~\protect{\ref{tab:exp}}. The values $\theta_{0,\infty}^\mc$ for $d=2$ and $d=3$ indicated here are reported, respectively, as gray and black vertical bars for $D=1$ in 
Figs.~\protect{\ref{fig:Gaux} and~\protect{\ref{fig:beyondGaux}}}. 
}
\label{tab:expMC}
\end{table}
As far as $\theta_0$ for $D=1$ is concerned [see Fig.~\ref{fig:beyondGaux}(b)], we note that the preliminary Monte Carlo estimates of Ref.~\cite{manifold} (black and grey circles, corresponding to $\theta_0^{\mc}(d=3,d'=2) \simeq 0.88$ and $\theta_0^{\mc}(d=2,d'=1) \simeq 0.72$, respectively) are essentially compatible with the values of  
$\theta_0^\bg$ [dash-dotted line in Fig.~\ref{fig:beyondGaux}(b); see also Tab.~\ref{tab:expMC}] both in $d=2$ (grey) and $d=3$ (black). The Monte Carlo estimates  $\theta_0^\mc$ discussed in Secs.~\ref{sec:MC3d} and~\ref{sec:MC2d} and reported in Tab.~\ref{tab:expMC}, instead, turn out to be significantly larger than and marginally compatible with the previous ones in $d=3$ and $d=2$, respectively, and therefore they no longer agree with  the corresponding analytical predictions provided by $\theta_0^\bg$.  
The reason for such a discrepancy might be that for $D=1$ the actual contribution of the term of ${\cal O}(D)$ in $\mu_0\times{\mathcal R}_0^\bg$ [see Eq.~\reff{eq:Reps}] is a rather large fraction of the ${\cal O}(D^0)$ term (ca. $100\%$ and $150\%$ in $d=3$ and $d=2$, respectively), \ie, the dependence on $D$ is so pronounced that a quantitatively reliable estimate  of these exponents might require accounting for higher-order terms in the codimension expansion. 

\smallskip
We conclude this section by discussing briefly the case $0 \leq \zeta \leq
1$ (see Tab.~\ref{fig_table}), that includes the instance of a
line $d'=1$ within the three-dimensional Ising model (\ie, $D=2$) for which $\zeta = \eta/z
\simeq 0.016$ (see Tab.~\ref{tab:exp}). 
For $0\le \zeta \le 1$, our analysis indicates that the long-time behavior of
$P_c(t)$ is independent of the actual value $m_0$ (\eg, $m_0=0$ or $m_0 = 1$) 
and is characterized by a stretched exponential behavior
(Tab.~\ref{fig_table}). 
Unfortunately, in our simulations, we have not
been able to observe incontrovertibly such a stretched-exponential
law. This difficulty mirrors the one recently encountered in the 
numerical analysis of the persistence probability $P_c(t)$ of stationary processes
characterized by two-time correlations with a power-law decay
$C_{\rm st}(t_1 , t _2) \sim |t_1-t_2|^{-\zeta}$~\cite{bunde_stretched,moloney}. 
Although that context is rather different from the present, there it was observed that the convergence to the stretched exponential behavior --- which is 
expected on the basis of the theorem by Newell and Rosenblatt~\cite{newell} ---
is actually extremely slow. In addition, the pre-asymptotic behavior was shown 
to be increasingly important at the quantitative level as $\zeta$ decreases~\cite{moloney}. 
Given these results and the extremely
small value of $\zeta = 0.016$ in the case of present interest it is not surprising
that the stretched exponential is rather difficult to be observed. 
On the other hand, our numerical data indicate that the pre-asymptotic
behavior of $P_c(t)$ is affected by the actual value of the initial
magnetization $m_0$. However, beyond this qualitative feature, we have not attempted 
a more detailed and quantitative characterization of this pre-asymptotic regime, which would 
require more extensive and dedicated simulations.

\section{$O(n)$ model in the limit $n \to \infty$}

\label{sec:On}

Here we extend the previous analysis of the persistence probability of a manifold to the case of a $n$-component vector order parameter 
$\varphi = (\varphi_1, \varphi_2, \cdots, \varphi_n)$ with $n > 1$ and model A dynamics [see Eqs.~\reff{def_Langevin} and~\reff{def_O1}]. 
Even though, strictly speaking, this model is not relevant for the description of actual physical systems, the fact that it can be  solved exactly in the limit $n\rightarrow \infty$ provides insight into the effects of non-linear terms in the Langevin equation beyond perturbation theory.
If the initial magnetization $m_0$ does not vanish the original $O(n)$ symmetry of the model is explicitly broken and one has to distinguish between fluctuations $\psi_\sigma(\bx,t)$ which are parallel  to the average order
parameter $m(t) = \langle \varphi(\bx, t) \rangle$ and those  $\psi_\pi(\bx,t)$ which are transverse to it.  The different fluctuations $\psi_\sigma(\bx,t)$ and $\psi_\pi(\bx,t)$ are expected to have
distinct persistence probabilities $P_c^\sigma(t)$ and $P_c^\pi(t)$ which, on the basis of the results of Ref.~\cite{andrea_ordered_on}, can be shown to exhibit the same temporal crossovers as in Tab.~\ref{fig_table} due to
$m_0\neq 0$. 

\subsection{Persistence of the transverse modes in the $O(n\rightarrow\infty)$ model with finite $m_0$}

\label{sec:Onper}

We focus here on the limit $n \to \infty$ and we consider the transverse modes only, for which the response and correlation functions can be calculated exactly~\cite{andrea_ordered_on}. The wave-vector dependent response function is given by (see Eq.~(68) in Ref.~\cite{andrea_ordered_on})
\be
R^\pi_{\bQ}(t>t',t') \equiv \frac{\delta \langle \psi^\pi(\bQ,t)\rangle}{\delta h^\pi(-\bQ,t')} \Bigg |_{h^\pi=0} =  \rme^{-Q^2 (t-s)}\left(\frac{t}{s}\right)^{\eps/4}
\left(\frac{1+s/\tau_m}{1+t/\tau_m}\right)^{1/2} \;,
\label{eq:Rpi}
\ee
from which one obtains the correlation function (see Eq.~(56) in Ref. \cite{andrea_ordered_on}) as
\be
\begin{split}
C^\pi_{\bQ}(t,t') &\equiv \langle \psi(\bQ,t) \psi(-\bQ,t') \rangle  = 2 \int_0^{t_<} \rmd t_1 R^\pi_{\bQ}(t,t_1)R^\pi_{\bQ}(t',t_1)\\
&=\frac{2(tt')^{\eps/4}}{\left[(1+t/\tau_m)(1+t'/\tau_m)\right]^{1/2}}\rme^{-Q^2(t+t')}
  \int_0^{t_<} \rmd t_1 (t_1)^{-\eps/2}(1+t_1/\tau_m)\rme^{2Q^2 t_1} \;,
\end{split}
\label{eq:Cpi}
\ee
where $\eps = 4-d$. Due to the residual $O(n-1)$ symmetry in the internal space of the transverse components, only correlation functions between the same component of the vector $\psi^\pi$ can be non-vanishing. Accordingly, here and in the following we always refer to such correlations even if this is not explicitly indicated. For $m_0\rightarrow\infty$, $\tau_m\sim m_0^{-2} \rightarrow 0$ and therefore one has 
\be
\begin{split}
R^\pi_{\bQ}(t>t',t') &= \rme^{-Q^2 (t-t')}\left(\frac{t}{t'}\right)^{\eps/4-1/2}
\quad \mbox{and}\\
C^\pi_{\bQ}(t_1,t_2) &= 2 (t_1t_2)^{\eps/4-1/2} \rme^{-Q^2(t_1+t_2)}
\int_0^{t_<}\rmd t' \rme^{2Q^2 t'} (t')^{1-\eps/2} \;.
\end{split}
\ee
Accordingly, the connected correlation function of the magnetization $\tm^\pi(\br,t)$ of a
manifold  defined as in Eq.~\reff{def_mag_tilde} with the substitution $\psi \to \psi^\pi$ is
\be
\begin{split}
C^\pi_\tm(t,t') =& \langle  \tm^\pi(\br,t) \tm^\pi(\br,t') \rangle = \int \prod_{i=1}^{D} \frac{\rmd q_i}{2 \pi} C^\pi_{{\bf Q}=(0,\bq)}(t,t')\\
=& 2 c_D2^{-D/2} (tt')^{\eps/4-1/2} {t'}^{2-(\eps+D)/2} \int_0^1\rmd u\;
u^{1-\eps/2} (y-u)^{-D/2} \;,
\end{split}
\label{eq:Xpi}
\ee
where $t > t'$, $y \equiv (1 + x^{-1})/2$, and $x=t'/t < 1$ [$c_D$ was defined after Eq.~\reff{eq:cmt}]. The normalized process $X^\pi(\br,t) = \tm^\pi(\br,t)/\langle [\tm^\pi(\br,t)]^2 \rangle^{1/2}$ is therefore characterized by the correlation function
\begin{equation}
\langle X^\pi(\br, t)X^\pi(\br, t')\rangle =(t/t')^{-(d+D)/4} {\mathcal A}^\pi(t'/t;d) \;,
\label{eq:corrXpi}
\end{equation}
which is of the form~\reff{scaling_x} with $F_\infty(t/t') = {\mathcal A}^\pi(t'/t;d)$, where
\begin{equation}
 {\mathcal A}^\pi(x;d) = x^{-D/2}\frac{\int_0^1\rmd u\; u^{-1+d/2} (y -
 u)^{-D/2}}{\int_0^1\rmd u\; u^{-1+d/2} (1 - u)^{-D/2}} \;,
\label{eq:defApi}
\end{equation}
(well defined for $d>0$ and $0<D<2$) and $\mu_\infty(d,d') = (d+D)/4$. This expression for $\mu_\infty$ agrees with  Eq.~\reff{eq:hyp_a} in which $\theta$ takes the value $\theta = -\beta/(\nu z) = -(d-2+\eta)/(2 z)$ characterizing the scaling form Eq.~\reff{scaling_form_manifold} for the transverse modes (see, \eg, Refs.~\cite{andrea_ordered_on,fedo}) and which yields, in general, $\mu_\infty = 1+[(d+D)/2-2+\eta]/z$. For this specific model, $\eta=0$ and $z=2$.
According to Eqs.~\reff{eq:corrXpi} and~\reff{eq:defApi}
the process $X^\pi$ is Markovian for vanishing codimension $D=0$, with 
$\bar\mu_\infty = d/4$. The corrections due to $D\neq 0$ can be accounted for
perturbatively by expanding $\bar{\cal A}^\pi(x;d) \equiv x^{D/4}{\cal A}^\pi(x;d)$ for small $D$ as
\begin{eqnarray}
&&\bar{\mathcal A}^\pi(x;d) = 1 + D \bar{\mathcal A}_1^\pi(x;d) + {\cal O}(D^2), \quad\mbox{with} \\
&& \bar{\mathcal A}_1^\pi(x;d) =  - \frac{1}{4}\ln x - \frac{d}{4} \int_0^1 \rmd u\;
  \left. u^{-1+d/2}\ln\frac{y-u}{1-u}\right|_{y=(1+x^{-1})/2} \;.
\end{eqnarray}
After some algebra one obtains from Eq.~\reff{theta_smallD},
\be
\bar{\mathcal R}_d \equiv \frac{\theta_\infty}{\bar\mu_\infty} = 1 + D I_d + {\cal O}(D^2) \;,
\ee
where
\be
I_d \equiv -\frac{1}{d} +\frac{d^2}{8\pi}
\int_0^1\rmd x \frac{x^{-1+d/4}}{(1-x^{d/2})^{3/2}}  \int_0^1 \rmd u\;
  \left. u^{-1+d/2}\ln\frac{y-u}{1-u}\right|_{y=(1+x^{-1})/2} 
\ee 
is a monotonically decreasing function of $d$, with 
$I_1=1.149\ldots$, $I_2 = \sqrt{2} -1/2=0.9142\ldots$, $I_3 = 0.867\ldots$, and $I_4 = 1/\pi + \pi/4 - 1/4=
0.853\ldots$.  For later convenience we indicate here also the values of $I_d$ for some $d\le 2$.
This provides the estimate
\begin{equation}
\theta_\infty^{[1]} = \frac{d}{4} [1 + D\times I_d + {\cal O}(D^2)] \;,
\label{eq:thinfsp1}
\end{equation}
which, for $D=0$, renders the value of the global persistence exponent reported  for this model in Ref.~\cite{us_epl}.
As discussed in Sec.~\ref{sec:analyt}, a different estimate of $\theta_\infty$ can be obtained by expanding only the ratio ${\mathcal R}_d$ to first order in the codimension $D$, while keeping the full $D$-dependence of the Markovian exponent $\mu_\infty = (d+D)/4$, which corresponds to ${\mathcal A}^\pi \equiv 1$ in Eq.~\reff{eq:corrXpi} and to $\bar {\mathcal A}_1^\pi(x;d) \mapsto {\cal A}^\pi_1(x;d) \equiv \bar{\mathcal A}^\pi_1(x;d) - (\ln x)/4$ in Eq.~\reff{theta_smallD}. This yields $\theta^{[2]}_\infty = \mu_\infty\times {\mathcal R}$, with ${\mathcal R} = \bar{\mathcal R} - D/d + {\cal O}(D^2)$, \ie,
\begin{equation}
\theta^{[2]}_\infty = \frac{d+D}{4} [1 + D \times (I_d - 1/d)+ {\cal O}(D^2)] \;.
\label{eq:thinfsp2}
\end{equation}
As expected, $\theta^{[2]}_\infty$ has the same small-$D$ expansion as $\theta^{[1]}_\infty$ up to ${\cal O}(D^2)$.

For completeness and comparison we briefly present here also the expansion for $\theta_0$ within the same $O(n\rightarrow\infty)$ model, which was first discussed in Ref.~\cite{manifold}. As the initial value $m_0$ of the magnetization vanishes, the original $O(n)$ symmetry is restored and there is no longer distinction between the transverse ($\pi$) and longitudinal ($\sigma$) fluctuation modes. The correlation function $\langle X(\br,t)X(\br,t')\rangle$ of the normalized process associated to the magnetization $\tm(\br,t)$ of the manifold can be obtained from Eq.~\reff{eq:Xpi}, with $ C^\pi_{{\bf Q}=(0,\bq)}(t,t')$ given by the limit for $\tau_m \rightarrow\infty$ of Eq.~\reff{eq:Cpi}:
\begin{equation}
C^\pi_{\bQ}(t,t') = 2(tt')^{\eps/4} \rme^{-Q^2(t+t')}
  \int_0^{t_<} \rmd t_1 (t_1)^{-\eps/2}\rme^{2Q^2 t_1}.
\end{equation}
Interestingly enough, this expression of $C^\pi_{\bQ}(t,t')$ with $\tau_m=\infty$ [and, analogously, the expression for $R^\pi_{\bQ}(t,t')$, see Eq.~\reff{eq:Rpi}] can be formally obtained from the one corresponding to $\tau_m=0$ [see Eq.~\reff{eq:Xpi}] by substituting in the latter $\eps$  with $\eps+2$, \ie, $d$ with $d-2$. Accordingly, we can take advantage here of the results reported above for the case $\tau_m=0$ in order to discuss the persistence properties of the manifold for $m_0=0$.
In particular, for the normalized process $X(\br,t) = \tm(\br,t)/\langle [\tm(\br,t)]^2 \rangle^{1/2}$, one finds
\begin{equation}
\begin{split}
\langle X(\br, t)X(\br, t')\rangle  &= \left.\langle X^\pi(\br, t)X^\pi(\br, t')\rangle \right|_{d\mapsto d-2}\\
& = (t/t')^{-(d+D-2)/4} {\mathcal A}^\pi(t'/t;d-2),
\end{split}
\label{eq:corrXpi0}
\end{equation}
on the basis of Eq.~\reff{eq:corrXpi}.
This result is of the form~\reff{scaling_x} 
with $\mu_0(d,d') = (d+D-2)/4$ and $F_0(t/t') = {\mathcal A}^\pi(t'/t;d-2)$.
As expected, for $d=4$ the equations above render  the corresponding ones for the Gaussian model with $m_0=0$, which were discussed at the end of Sec.~\ref{sec:Gaux}. In addition, this expression for $\mu_0$ agrees with the hyperscaling relation [analogous to Eq.~\reff{eq:hyp_b}] briefly mentioned after Eq.~\reff{scaling_gaussian_0} because $\eta=0$ and $\theta = (4-d)/4$ for the present $O(n\rightarrow\infty)$ model~\cite{janssen_rg}.
On the basis of the mapping highlighted above, one can take advantage of Eqs.~\reff{eq:thinfsp1} and~\reff{eq:thinfsp2} in order to determine the expansion of $\theta_0(d,d-D)$ in the codimension $D$:
\begin{equation}
\theta_0^{[1]} = \frac{d-2}{4} [1 + D\times I_{d-2} + {\cal O}(D^2)],
\label{eq:th0sp1}
\end{equation}
(which reproduces the expansion provided in Ref.~\cite{manifold} for $d=4$, whereas for $d=3$ this gives a coefficient $I_1/4 =0.287\ldots$ which corrects the value $0.183615\ldots$ reported therein for the first-order correction in $D$) 
and
\begin{equation}
\theta^{[2]}_0 = \frac{d-2+D}{4} \{1 + D \times [I_{d-2} - 1/(d-2)]+ {\cal O}(D^2)\}\,.
\label{eq:th0sp2}
\end{equation}
%
%
%
%
\begin{figure}
\begin{tabular}{ccc}
\begin{minipage}{0.4\linewidth}
\includegraphics[scale=0.6]{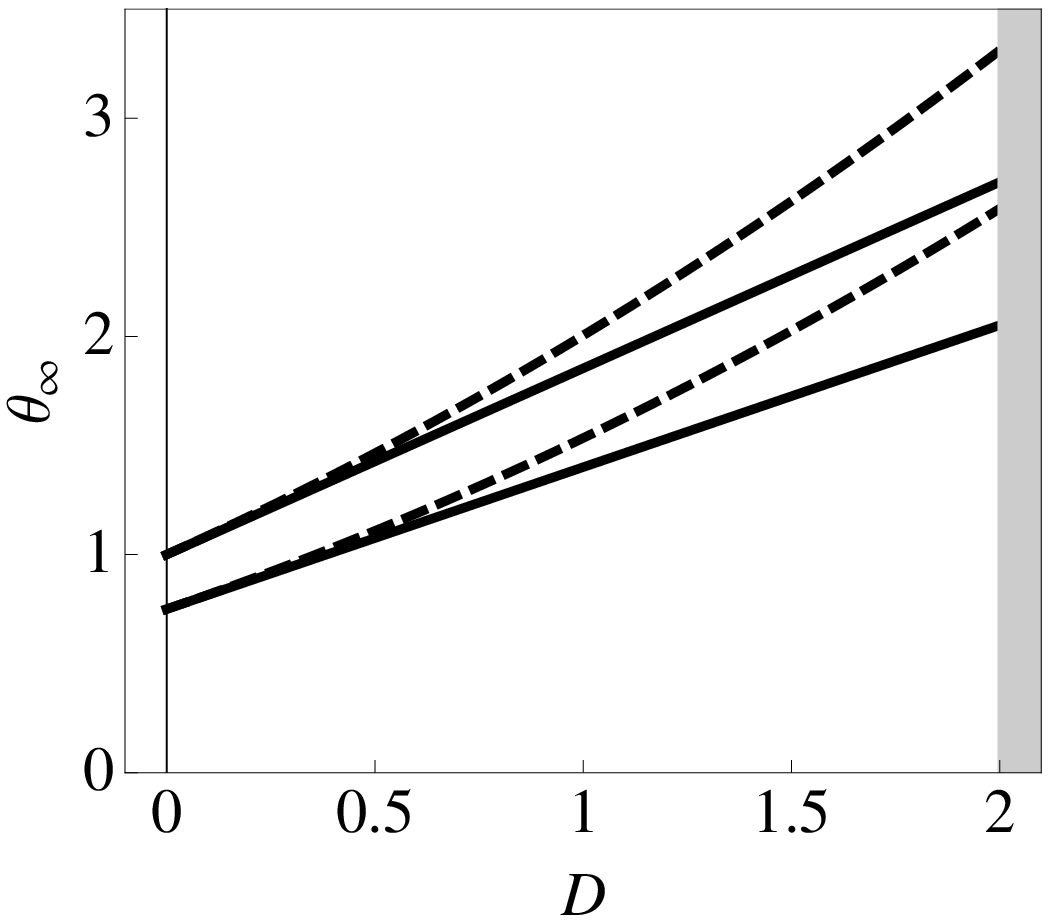}
\end{minipage}
&\quad\quad\quad&
\begin{minipage}{0.4\linewidth}
\vspace{-1mm}

\includegraphics[scale=0.6]{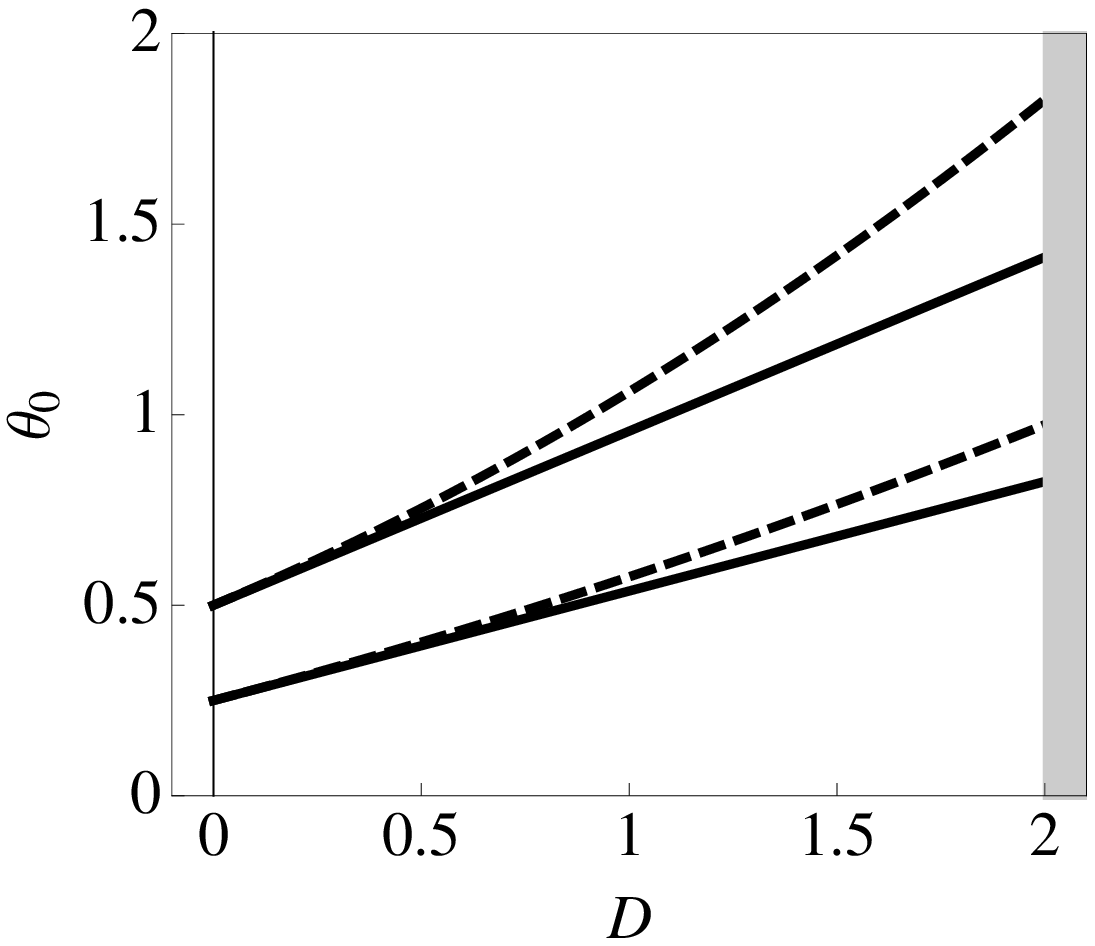}
\end{minipage}\\
(a) && (b)
\end{tabular}
\caption{Persistence exponents  $\theta_{\infty,0}(d,d-D)$ of  the transverse components of the global order parameter of a manifold of codimension $D$ as a function of $D$, for $d=4$ (uppermost solid and dashed curves) and $d=3$ (lowermost solid and dashed curves),  within the $O(n\rightarrow\infty)$ model universality class with relaxational dynamics [Eq.~\reff{def_Langevin}].
Panels (a) and (b) refer to the case of non-vanishing and vanishing initial value $m_0$ of the order parameter, respectively, whereas 
the two different estimates $\theta^{[1]}_{\infty,0}$  [Eqs.~\reff{eq:thinfsp1} and~\reff{eq:th0sp1}] and $\theta^{[2]}_{\infty,0}$ [Eqs.~\reff{eq:thinfsp2} and~\reff{eq:th0sp2}] are indicated by the  the solid and the dashed curves, respectively.
For $D\ge 2$, \ie, $\zeta\ge0$ (shaded areas) the asymptotic decay of the persistence probability of this model is no longer algebraic. In panel (a), the uppermost solid and dashed curves ($d=4$) are the same as those reported in Figs.~\protect{\ref{fig:Gaux}}(b) and \protect{\ref{fig:beyondGaux}}(b). 
The persistence exponent of the longitudinal component of the global order parameter for $m_0=0$  is the same as the one for the transverse component [panel (b)], whereas for $m_0\neq 0$ and $d=4$ (Gaussian approximation) it is given by the solid and dashed lines in Figs.~\protect{\ref{fig:Gaux}}(a) and \protect{\ref{fig:beyondGaux}}(a). 
The case of the longitudinal component for $m_0=0$ was also discussed in Ref.~\cite{manifold}.
\label{fig:On}}
\end{figure}
%
%
In Fig.~\ref{fig:On}(a) and (b) we report the estimates $\theta^{[1,2]}_\infty$  for $\theta_\infty$ [Eqs.~\reff{eq:thinfsp1} and~\reff{eq:thinfsp2}] and $\theta^{[1,2]}_0$  for $\theta_0$  [Eqs.~\reff{eq:th0sp1} and~\reff{eq:th0sp2}, see also Ref.~\cite{manifold}], respectively, as functions of the codimension $D$. In the two panels $\theta^{[1]}_{\infty,0}$ and $\theta^{[2]}_{\infty,0}$ are indicated as solid and dashed curves, respectively, for $d=4$ (uppermost solid and dashed curves) and $d=3$ (lowermost solid and dashed curves).  Comparing the curves for $d=3$ to those corresponding to $d=4$ (Gaussian approximation), it turns out that the effect of non-Gaussian terms in the effective Hamitonian~\reff{def_O1} for $n\rightarrow\infty$ is to reduce the value of both $\theta_\infty$ and $\theta_0$ for generic values of $D$. This trend agrees with the one observed at the end of Sec.~\ref{sec:BGaux} both for the longitudinal and the transverse components of the order parameter and based on the results of a dimensional expansion around $d=4$ for the case $D=0$ and generic $n$.
Even though a definitive statement in this respect would require a careful analysis, one might heuristically expect on the basis of the evidences collected here that, for a fixed codimension $D\neq 0$, the value of the persistence exponents $\theta_{\infty,0}$ \emph{decreases} as $d$ decreases below the upper critical dimensionality $d=4$ of the $O(n\to\infty)$ model, as it happens for $D=0$. 

\subsection{Relation between the spherical model and the $O(n\rightarrow \infty)$ model}
\label{sec:sp-On}

We conclude this section by discussing the relation between the
$O(n\rightarrow \infty)$ model studied in Sec.~\ref{sec:Onper} and the spherical model (see, \eg, Ref.~\cite{Joyce}),
and its implications for the persistence probability.
It is well-known that these two models are equivalent in equilibrium as
there is a mapping between the corresponding free energies. 
This equivalence also extends to their equilibrium dynamics. 
However, when considering non-equilibrium properties some aspects of the spherical model 
require a careful consideration. In particular, non-Gaussian fluctuations of the Lagrange multiplier which is
introduced in order to impose the pure relaxational dynamics have to be
accounted for when calculating some correlation functions of \emph{global} 
quantities --- which correspond to vanishing wave-vector ${\bQ}=0$ (see Refs.~\cite{as-06,as-new}). 
In doing so, it turns
out that the behavior of these quantities cannot be obtained as the limit for
${\bQ}\rightarrow 0$ of the correlation functions for ${\bQ}\neq 0$
(\emph{local}), given that the associated non-connected part (if non-vanishing)
alters the scaling behavior compared to the case ${\bQ}\neq 0$. 
(Here and in what follows we assume that the system is spatially homogeneous.)

Consider, for example, the correlation function of the local order parameter (\ie, of the magnetization). 
As long as the average of the local magnetization vanishes --- which is the case if $m_0 = 0$ --- 
there are no differences between the connected and the non-connected correlation functions. 
However, as soon as the
average of the local magnetization is non-zero (\eg, when the system is prepared in a
magnetized initial state $m_0\neq 0$), the connected correlation
function of the magnetization differs from the non-connected one only at
${\bf Q} = 0$ and the subtraction of the non-connected part alters its original 
scaling behavior. In turn, this subtraction affects also the scaling behavior of the
response function.
For a different observable, \eg, the energy, this
difference in the scaling behavior might emerge also in the case of $m_0=0$.
Focusing here on the correlation and response functions of the order parameter of
the spherical model ($S$), one finds that for $m_0 =0$, the correlation and response
function of global quantities can be simply obtained as the limit ${\bf
  Q}\rightarrow 0$ of the local quantities corresponding to a non-vanishing
${\bf Q}$ and that they coincide with the same quantities in the
$O(n\rightarrow\infty)$ model. 
For $m_0\neq 0$ and  \emph{local} space integrals of the order parameter (\ie, ${\bf
  Q}\neq 0$, for which the non-connected part vanishes), instead,
the correlation and response functions are given by (see Eqs.~(2), (13) and
  (22) in Ref.~\cite{as-new})
\be %
R^S_\bQ(t>t',t') =
  \left(\frac{t}{t'}\right)^{\rho/2}\left(\frac{1+t'/\tau_m}{1+t/\tau_m}\right)^{1/2}
  \rme^{-\omega (t-t')} \;,
\label{eq:Rlocsp}
\ee
where $\omega \equiv \omega_\bQ = 2 \sum_{i=1}^d(1-\cos Q_i)$, $\rho =
  (4-d)/2 = \eps/2$ for $2<d<4$, whereas $\rho=0$ for $d>4$ 
(see Eq.~(14) in Ref.~\cite{as-new}),  
$\tau_m = c m_0^{-2}$ 
(see Eq.~(15) in Ref.~\cite{as-new}). The
  correlation function turns out to be, for $t>t'$
\be
C^S_\bQ(t,t') = T_c \times  \frac{2(tt')^{\rho/2}}{\left[(1+t/\tau_m)(1+t'/\tau_m)\right]^{1/2}}\rme^{-\omega(t+t')}
  \int_0^{t_<} \rmd t_1 (t_1)^{-\rho}(1+t_1/\tau_m)\rme^{2\omega t_1} \;.
\label{eq:Clocsp}
\ee
In Eqs.~\reff{eq:Rlocsp} and \reff{eq:Clocsp}, the superscript $S$ indicates that the expression refers to the spherical model. Comparing Eqs.~\reff{eq:Rlocsp} and~\reff{eq:Clocsp} with Eqs.~\reff{eq:Rpi} and \reff{eq:Cpi},
respectively, one concludes that, in the limit of small momenta (such that
$\omega = \omega_\bQ \simeq Q^2$),
\be
R^S_{\bQ\neq 0}(t,t') = R^\pi_\bQ(t,t')\quad\mbox{and}\quad C^S_{\bQ \neq 0}(t,t') =
C^\pi_\bQ(t,t') \;,
\label{eq:eqpi}
\ee
up to non-universal factors, for all dimensions $d>2$ (one can easily
check that the equality holds also in  the case $d>4$, for which the Gaussian approximation
is exact), and for all values of the initial magnetization $m_0\neq 0$. We
emphasize that the left-hand sides of Eq.~\reff{eq:eqpi} refer to the order parameter of the spherical model, whereas 
the corresponding right-hand sides to the \emph{transverse} fluctuations ($\pi$) of the order parameter of the 
$O(n\rightarrow\infty)$ model. These equalities justify the fact --- already
pointed out right after Eqs.~(73) and (126) in Ref.~\cite{andrea_ordered_on}  and at the
end of Sec.~5  in Ref.~\cite{as-new} --- that the asymptotic value of the
fluctuation-dissipation ratio for \emph{local} integrals of the order
parameter in the spherical model is the same as the corresponding one for \emph{transverse} fluctuations
in the $O(n\rightarrow\infty)$ model. 

Consider now the \emph{global} order parameter, \ie, the
integral over the whole space of the local order parameter, corresponding to
the Fourier mode with ${\bf Q}= 0$: its statistical average does not vanish for
$m_0\neq 0$ and therefore its connected two-time correlation function differs
from the non-connected one and cannot be simply obtained as the limit for
${\bf Q}\rightarrow 0$ of the same quantity for ${\bf Q} \neq 0$. The explicit
expression for the correlation $C_{\bQ=0}$ and response $R_{\bQ=0}$ functions of
the global order parameter for $d>4$ and arbitrary value of the magnetization $m_0$
were reported in Ref.~\cite{as-new} [see Eqs.~(48) and~(47) therein]:
\be
R^S_{\bQ=0}(t>t',t') = \left(\frac{1+t'/\tau_m}{1+t/\tau_m}\right)^{3/2} \;,
\label{eq:gRsph}
\ee
and (written in a slightly different form compared to Ref.~\cite{as-new})
\be
C^S_{\bQ=0}(t,t') = T_c\times \frac{2\int_0^{t_<}\rmd t_1
  (1+t_1/\tau_m)^3}{\left[(1+t/\tau_m)(1+t'/\tau_m)\right]^{3/2}} \;.
\label{eq:gCsph}
\ee
By comparing these expressions with Eqs.~\reff{eq:Rlocsp}
and~\reff{eq:Clocsp}, one realizes that --- as anticipated above --- 
the former \emph{are not} given by the limit $\bQ\rightarrow 0$ (\ie,
$\omega \rightarrow 0$) of the latter. On the other hand, the expressions for
$R^S_{\bQ=0}$ and $C^S_{\bQ=0}$ for the spherical model in $d>4$ are the same (up to an
irrelevant multiplicative factor for $C$) as the one for the
\emph{longitudinal} 
fluctuations ($\sigma$) with ${\bf Q}=0$ of the order parameter  of the $O(n\rightarrow\infty)$ model, given by  Eqs.~(58)
and~(59) of Ref.~\cite{cgk-06}:
\be
R^S_{\bQ=0}(t,t') = R^\sigma_{\bQ=0}(t,t')\quad\mbox{and}\quad C^S_{\bQ=0}(t,t') =
C^\sigma_{\bQ=0}(t,t') \quad\quad\quad (d>4) \;,
\label{eq:eqsigma}
\ee
where, as the superscripts indicate, the left-hand sides refer to the spherical model whereas the right-hand sides to the
longitudinal fluctuations in the $O(n\rightarrow\infty)$ model.
(Note
that for the case $d>4$ presently discussed there is no difference between the
longitudinal fluctuations of the $O(n)$ and $O(1)$ model, and therefore the
expressions for $R^\sigma_{\bQ=0}$ and $C^\sigma_{\bQ=0}$ 
can be read from Ref.~\cite{cgk-06}, in which the
latter, \ie,the Ising model is actually studied.)
Even though we have proven the relation~\reff{eq:eqsigma} only in the case $d>4$, one
heuristically expects it to be valid also for $d<4$, as it was the case for
the analogous relation~\reff{eq:eqpi}. The expressions for the correlation and
response function of the global magnetization of the spherical model for $2<d<4$ and a generic value
of the initial magnetization $m_0$ (\ie, $\tau_m$) can be
found in Ref.~\cite{as-new}. While the expression for $R^S_{\bQ=0}$ is explicitly
given by Eq.~(58) therein,
\be
R^S_{\bQ=0}(t>t',t') = \left(\frac{t}{t'}\right)^{1-d/4}
\left(\frac{1+t'/\tau_m}{1+t/\tau_m}\right)^{1/2} 
\left[1 - \frac{t/\tau_m}{1+t/\tau_m}\left(1- \frac{t'}{t}\right)^{d/2-1}
  \right] \;,
\label{eq:Rglosph}
\ee
the corresponding expression for the correlation function $C^S_{\bQ=0}(t,t')$ is significantly
more involved and it has been worked out explicitly only in the asymptotic
regime $t' \ll t$. For the $O(n)$ model, instead, the response and
correlation functions $R^\sigma_{\bQ=0}(t,t')$ and $C^\sigma_{\bQ=0}(t,t')$, respectively, 
have been calculated for $d<4$ only at the first order
in the $\eps$-expansion around $d=4$ (with $\eps = 4-d$) and in the limit of
large magnetization, \ie, for $\tau_m \rightarrow 0$. These expressions,
reported in Ref.~\cite{andrea_ordered_on} can be compared in the limit $n\rightarrow\infty$ 
with the first term of the expansion around $d=4-\eps$ of the
corresponding result for the spherical model. In particular, focusing on the
response function for $n\rightarrow\infty$, one has (see Eqs.~(88) and (89) in
Ref.~\cite{andrea_ordered_on})
\be
\begin{split}
R^\sigma_{\bQ=0}(t>t',t') &= \left(\frac{t'}{t}\right)^{3/2}\left\{1 - \frac{\eps}{4}
\ln \frac{t'}{t} +
\frac{\eps}{2}\left(\frac{t}{t'}-1\right)\ln\left(1-\frac{t'}{t}\right)\right\}+
O(\eps^2) \\
&= \left(\frac{t'}{t}\right)^{(2-\eps)/4} \left\{ 1 - \left(
1-\frac{t'}{t}\right)^{1-\eps/2}\right\} + O(\eps^2) \;,
\end{split}
\ee
which is indeed equal to the first-order expansion of Eq.~\reff{eq:Rglosph} around $d=4$ and for
$\tau_m= 0$.
In order to do an analogous comparison
for the correlation functions $C^S_{\bQ=0}(t,t')$ and
$C^\sigma_{\bQ=0}(t,t')$ for $d<4$ one would have to calculate the limit $n\rightarrow\infty$
of the results for $C^\sigma_{\bQ=0}(t,t')$ 
presented in Ref.~\cite{andrea_ordered_on} for $\tau_m=0$ [in particular, see Eqs.~(93), (B31), (B24),
and (B29) therein] and compare it with the limit $\tau_m\rightarrow 0$ of the
correlation function $C^S_{\bQ=0}(t,t')$ of the spherical model explicitly presented in
Ref.~\cite{as-new} for the case $t'\ll t$. However, this somewhat lengthy calculation can be avoided
by noting that this equality up to ${\cal O}(\eps^2)$ is implied by the one between
the response functions of the spherical and of the $O(n\rightarrow\infty)$
model up to ${\cal O}(\eps^2)$, together with the fact that the asymptotic value of
the (fluctuation-dissipation) ratio $[\partial_{t'} C(t,t')]/R(t,t')$ for $t\gg t'$ and $\tau_m
\rightarrow 0$ is the same in the two models [see right after Eq.~(105) in Ref.~\cite{as-new}].

Summing up, the results of
Refs.~\cite{as-06,as-new,cgk-06,andrea_ordered_on} suggest the equalities
Eqs.~\reff{eq:eqpi} and \reff{eq:eqsigma} for the order parameter response and correlation functions of the
spherical and of the $O(n\rightarrow\infty)$ models relaxing from an initial state with $m_0\neq 0$. 
Whereas Eq.~\reff{eq:eqpi} can be explicitly checked for $d>2$ and all values of the
initial magnetization $m_0\neq 0$, Eq.~\reff{eq:eqsigma} is proven for generic
values of $\tau_m$ only for $d>4$, while in the case $\tau_m=0$
also for $2<d<4$ but only in the limit $t'\ll t$. If, as it is likely, this
relation extends to the remaining cases, it would be interesting to understand its deeper motivation as it connects, together with Eq.~\reff{eq:eqpi}, different degrees of freedom in different models.

According to the correspondence highlighted above, for a spatially constant
non-vanishing initial value of the order parameter $m_0$, the
persistence properties of the \emph{global} order parameter of
the spherical model are (up to non-universal factors) the same as the one
of the global \emph{longitudinal} fluctuations $(\sigma)$ of the order parameter in the
$O(n\rightarrow\infty)$ model. However, the persistence properties of the
global order parameter of a \emph{manifold} of non-vanishing codimension $D\neq 0$ 
in the spherical model are (up to non-universal factors) the same as the ones of the
global {\it transverse}  fluctuations $(\pi)$ of the order parameter of the same manifold in the
$O(n\rightarrow\infty)$ model. A consequence of this correspondence is that
the small codimension expansion which typically allows for a perturbative access to 
the persistence exponent of the manifold ($D\neq 0$) within the spherical
model \emph{is not} an expansion about the persistence exponent of the global order
parameter of the entire system ($D=0$), as highlighted by the fact that the former and the latter
actually involve different degrees of freedom ($\pi$ and $\sigma$, respectively) in the corresponding 
$O(n\rightarrow\infty)$ model. 

In passing we mention that the result reported in Ref.~\cite{us_epl} for the global persistence exponent $\theta_\infty$ of the spherical model (\ie, with $D=0$ and $m_0\neq 0$) is incorrectly based on the expression of the (connected) correlation function $C_m(t,t')$ of the global magnetization for $t \gg t'$ reported in Eq.~(8.108) of Ref.~\cite{as-06}. Indeed, the knowledge of $C_m(t\gg t',t')$ gives access only to the associated exponent $\mu_\infty = d/4+1$ within the Markovian approximation (which is correctly reported in Ref.~\cite{us_epl}), whereas  $\theta_\infty\neq \mu_\infty$ is actually determined by the full functional form of $C_m(t,t')$ for generic values of $t$ and $t'$, which is not explicitly provided in Ref.~\cite{as-06}.  The correspondence discussed above conveniently provides some information on the global (\ie, $D=0$) $\theta_\infty$ for spherical model with $m_0\neq 0$ on the basis of the corresponding results for the $O(n\rightarrow\infty)$ model presented in Ref.~\cite{us_epl}: Indeed, it implies that $\theta_\infty^S = \theta_\infty^\sigma$, in which the left-hand side refers to the spherical model whereas the right-hand side to the longitudinal fluctuations of the $O(n\rightarrow\infty)$ model.  For the $O(n)$ model one finds $\theta_\infty^\sigma = \mu_\infty^\sigma\times {\mathcal R}^\sigma$, where $\mu_\infty^\sigma = 1 + d/(2 z)$ and ${\mathcal R}^\sigma = 1 + \epsilon [(0.115\ldots + n 0.131\ldots)/(8+n)]+ {\cal O}(\epsilon^2)$ with $\epsilon = 4-d$~\cite{us_epl}, so that, in the limit $n\rightarrow\infty$, $\mu_\infty^\sigma \rightarrow  1 + d/4$ ($z\rightarrow 2$ \cite{HHM-72}) and ${\mathcal R}^\sigma \rightarrow 1 + \epsilon \times 0.131\ldots + {\cal O}(\epsilon^2)$. For the spherical model this implies that $\theta_\infty^S = (1 + d/4) \times[1 + \epsilon \times 0.131\ldots + {\cal O}(\epsilon^2)]$ which indeed shows a non-Markovian correction.  Even though it would be nice to have a direct check of this prediction on the basis of the results of Ref.~\cite{as-06} for generic $d$,  the direct calculation of $\theta_\infty^S$ turns out to be rather involved and it is beyond the scopes of the present study.

\section{Conclusions and perspectives}

\label{sec:conc}

In summary, we have investigated both analytically and numerically the persistence probability $P_c(t)$ of a $d'$-dimensional manifold within a $d$-dimensional system which relaxes at the critical point from an initial state with non-vanishing value of the magnetization or, more generally, order parameter $m_0$. Such a persistence probability is defined as the probability that the fluctuating order parameter does not cross its average value up to time $t$. 
Depending on the value of the parameter $\zeta \equiv (D-2+\eta)/z$, with $D=d-d'$ we found that, as in the case $m_0=0$, the long-time decay of $P_c(t)$ is (i) exponential for $\zeta > 1$, (ii) stretched exponential for $0 \leq \zeta \leq 1 $ and (iii) algebraic for $\zeta < 0$. While in the first two cases (i) and (ii) the asymptotic behavior of $P_c(t)$ is not affected by a finite value of $m_0$, in the third case $\zeta < 0$ we demonstrated that $P_c(t)$ exhibits a temporal crossover 
between an early-time and a distinct late-time algebraic decay, which are characterized by two different exponents $\theta_0(d,d')$ and $\theta_\infty(d,d')$, respectively, with $\theta_\infty(d,d') > \theta_0(d,d')$ (see Tab.~\ref{fig_table}). Analogously to the case $D=0$, the crossover is controlled by the time scale $\tau_m \propto m_0^{-1/\kappa}$.
The analytic determination of the associated exponents $\theta_{0,\infty}(d,d')$ is rather non-trivial already within the Gaussian approximation (which becomes exact for $d>4$) because the stochastic process under study turns out to be non-Markovian for $D\neq 0$. In order to calculate $\theta_\infty(d,d')$ ($\theta_0$ was already studied in Ref.~\cite{manifold}) we  performed a perturbative expansion up to the first order in the codimension $D=d-d'$ of the manifold [see Eqs.~\reff{eq:est1}, \reff{eq:est2}, \reff{eq:est1_0} and \reff{eq:est2_0}]. Then we presented a perturbative approach which allows one to calculate the effects of non-Gaussian fluctuations in a dimensional expansion around the space dimensionality $d=4$. Combining these two expansions we obtained analytic estimates of $\theta_\infty(d,d')$ and $\theta_0(d,d')$ up to order ${\mathcal O}(D, \epsilon)$ [see Sec.~\ref{sec:BGaux} and Fig.~\ref{fig:beyondGaux}]. 
In order to assess the reliability of these analytic estimates and  to go beyond the perturbation theory, we studied the critical relaxation of the Ising model with Glauber dynamics on a $d$-dimensional hypercubic lattice. In particular we computed the persistence probability of the magnetization of a line in $d=2$ and of a plane in $d=3$ --- corresponding to codimension $D=1$.  
In both these  cases we observed a temporal crossover in $P_c(t)$ between two algebraic decays,  as predicted by our analytical investigation and we determined the numerical estimates of the associated exponents $\theta_\infty^{\mc}(d=2,d'=1)$ and $\theta_\infty^{\mc}(d=3,d'=2)$ as well as more accurate estimates of $\theta_0^{\mc}(d=2,d'=1)$ and $\theta_0^{\mc}(d=3,d'=2)$. 
In the case of vanishing codimension $D=0$ --- primarily investigated in Ref.~\cite{us_epl} --- the agreement between the corresponding numerical and analytical estimates of $\theta_{0,\infty}$ are very good both in two and three dimensions, as summarized here in Fig.~\ref{fig:beyondGaux}(b). For $D=1$, instead, $\theta_\infty^{\mc}(d=3,d'=2)$ is in rather good agreement with the corresponding analytical estimate, whereas  $\theta_\infty^{\mc}(d=2,d'=1)$ is significantly larger than the value predicted by our perturbative calculation --- see Fig.~\ref{fig:beyondGaux}(a) and Sec.~\ref{sec:MCth} for the comparison.  In addition, it turns out that $\theta_\infty^{MC}(d=3,d'=2) < \theta_\infty^{MC}(d=2,d'=1)$ while all the analytical approaches presented here (including the analysis of the $O(n\rightarrow\infty)$ model) suggest that the converse should be true.  
Our numerical simulations also unveiled a non-trivial scaling of the persistence probability with the characteristic time $\tau_m$, which remains to be understood. These two latter intriguing features  of the persistence probability of the manifold surely deserve further investigations beyond the preliminary one presented here.

Finally, we have complemented our analysis of the persistence properties by a thorough comparison between the non-equilibrium dynamics of the $O(n \rightarrow \infty)$ model and of the spherical model. While they are strictly equivalent as far as their equilibrium properties are concerned, we have pointed out that such an equivalence has to be carefully qualified  when discussing their non-equilibrium dynamics. In particular, in the case of the critical relaxation from an initial state with non-vanishing order parameter, an unexpected  connection emerges between the local order parameter of the spherical model and the transverse components of the order parameter in the $O(n\rightarrow\infty)$ model as well as between the global order parameter of the former and the longitudinal components of the latter --- see Sec.~\ref{sec:sp-On} for details. This connection is related to the fact --- pointed out in Refs.~\cite{as-06,as-new} --- that within the spherical model the correlation function of {\it global} quantities (corresponding to a vanishing wave-vector $\bQ = 0$) can not be obtained as the limit $\bQ \to 0$ of the associated {\it local} correlations (corresponding to $\bQ \neq 0$).  

In view of the results presented here, it would be certainly  interesting to analyze the consequences of a finite initial magnetization $m_0$ on other relevant properties which characterize the temporal evolution of the thermal fluctuations $\delta m(t)$ of the magnetization of a manifold. As an example, it was recently shown in Refs.~\cite{godreche_excursion, garcia} that the longest excursion $l_{\max}(t)$ between two successive zeros of a stochastic process up to time $t$ is an interesting quantity which characterizes the "history" of the stochastic process and the asymptotic behavior of which depends qualitatively on the value of the persistence exponent of the process being larger or smaller than a certain critical value $\theta_c$~\cite{godreche_excursion}.
Given that, for the systems studied here,  $\theta_0(d,d') < \theta_c < \theta_\infty(d,d')$ one expects an intriguing dynamical crossover in the growth of the average $\langle l_{\max}(t) \rangle$ \cite{godreche_excursion}, which certainly deserves further investigations.  

\begin{acknowledgments}
AG is supported by MIUR within the program 
``Incentivazione alla mobilit\`a di studiosi stranieri e italiani residenti
all'estero''. GS acknowledges the hospitality of the Max-Planck Institut f\"ur Metallforschung
in Stuttgart, where part of this work was done. We thank S.N. Majumdar for
helpful discussions. RP wishes to thank CAM for fruitful discussions.
\end{acknowledgments}


\end{document}